\pdfoutput=1
\documentclass[openany]{book}
\usepackage[backend=biber, maxbibnames=9,style=numeric-comp,sorting=none]{biblatex}
\usepackage{amsmath}
\usepackage{amsfonts}
\usepackage{amssymb}
\usepackage{bbold}
\usepackage{framed}
\usepackage{xfrac}
\usepackage{graphicx}
\usepackage{gensymb}
\usepackage{tabularx}
\usepackage{booktabs}
\usepackage{multirow}
\usepackage{xcolor}
\usepackage{xspace}
\usepackage{listings}
\usepackage{subcaption}
\usepackage[pscoord]{eso-pic}% The zero point of the coordinate systemis the lower left corner of the page (the default).
\usepackage[hang,flushmargin]{footmisc}
\usepackage{suffix}
\usepackage{titlesec}
\usepackage{fancyhdr}
\usepackage{chngcntr} % allows adjustment of counting of figures/tables/equations etc
\usepackage[left=1in,top=1in,bottom=1in, right=1in]{geometry}
\usepackage[blocks]{authblk}
\usepackage{csquotes}

\usepackage[bookmarks=false]{hyperref}

\counterwithout{figure}{chapter}
\counterwithout{table}{chapter}
\counterwithout{equation}{chapter}

\newcommand\BackgroundPic{%
\put(0,0){%
\parbox[b][1.2\paperheight]{\paperwidth}{%
\vfill
\centering
\includegraphics[bb= 0 0 980 893, width=\paperwidth,height=0.4\paperheight,%
keepaspectratio]{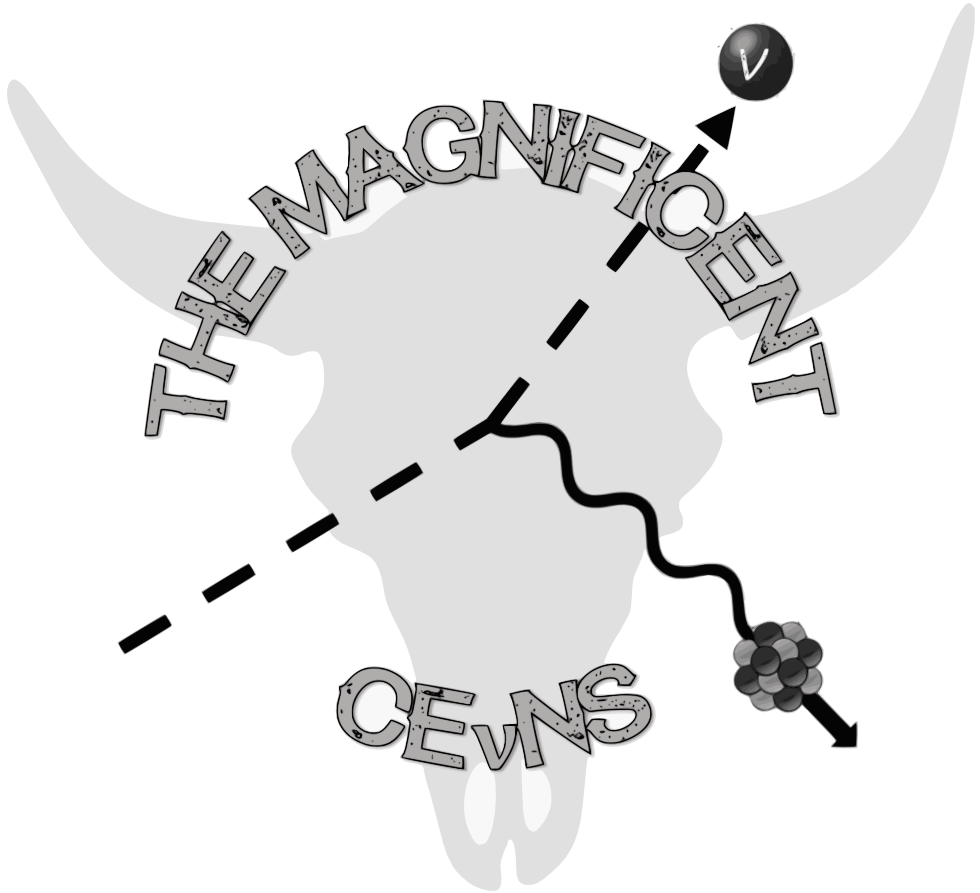}%
\vfill
}}}

\renewbibmacro{in:} {%
	\ifboolexpr{%
		test{\ifentrytype{article}}%
	}{}{\printtext{\bibstring{in}\intitlepunct}}%
}

\DeclareSourcemap{%
	\maps[datatype=bibtex]{
		\map[overwrite]{
			\step[fieldsource=shortjournal]
			\step[fieldset=journaltitle,origfieldval]
		}
		\map{
			\step[fieldsource=journal, match={Physics Letters},
				replace={Phys. Lett.}]
			\step[fieldsource=journal, match={Physics Letters A},
				replace={Phys. Lett. A}]
			\step[fieldsource=journal, match={Physics Letters B},
				replace={Phys. Lett. B}]
			\step[fieldsource=journal, match={Physical Review Letters},
				replace={Phys. Rev. Lett.}]
			\step[fieldsource=journal, match={Physical Review D},
				replace={Phys. Rev. D}]
			\step[fieldsource=journal, match={Physical Review C},
				replace={Phys. Rev. C}]
			\step[fieldsource=journal, match={Physical Review A},
				replace={Phys. Rev. A}]
			\step[fieldsource=journal, match={Physical Review X},
				replace={Phys. Rev. X}]
			\step[fieldsource=journal, match={Physical Review E},
				replace={Phys. Rev. E}]
			\step[fieldsource=journal, match={Nuclear Instruments and Methods in Physics Research Section A: Accelerators, Spectrometers, Detectors and Associated Equipment},
				replace={Nucl. Instrum. Meth. A}]
			\step[fieldsource=journal, match={Nuclear Instruments and Methods in Physics Research Section B: Beam Interactions with Materials and Atoms},
				replace={Nucl. Instrum. Meth. B}]
			\step[fieldsource=journal, match={Nuclear Instruments and Methods in Physics Research},
				replace={Nucl. Instrum. Methods}]
			\step[fieldsource=journal, match={Nuclear Instruments and Methods},
				replace={Nucl. Instrum. Methods}]
			\step[fieldsource=journal, match={The Astrophysical Journal Letters},
				replace={Astrophys. J. Lett.}]
			\step[fieldsource=journal, match={The Astrophysical Journal},
				replace={Astrophys. J.}]
			\step[fieldsource=journal, match={The Astrophysical Journal Supplement Series},
				replace={Astrophys. J. Suppl. S.}]
			\step[fieldsource=journal, match={Journal of Cosmology and Astroparticle Physics},
				replace={J. Cosmol. Astropart. P.}]
			\step[fieldsource=journal, match={Astroparticle Physics},
				replace={Astropart. Phys.}]
			\step[fieldsource=journal, match={International Journal of Modern Physics A},
				replace={Int. J. Mod. Phys. A}]
			\step[fieldsource=journal, match={International Journal of Modern Physics E},
				replace={Int. J. Mod. Phys. E}]
			\step[fieldsource=journal, match={Journal of Instrumentation},
				replace={J. Instrum.}]
			\step[fieldsource=journal, match={Journal of Physics G: Nuclear and Particle Physics},
				replace={J. Phys. G Nucl. Partic.}]
			\step[fieldsource=journal, match={Journal of Physics: Conference Series},
				replace={J. Phys. Conf. Ser.}]
			\step[fieldsource=booktitle, match={Journal of Physics: Conference Series},
				replace={J. Phys. Conf. Ser.}]
			\step[fieldsource=journal, match={Physics in Perspective},
				replace={Phys. Perspect.}]
			\step[fieldsource=journal, match={Progress in Particle and Nuclear Physics},
				replace={Prog. Part. Nucl. Phys.}]
			\step[fieldsource=journal, match={Journal of High Energy Physics},
				replace={J. High Energy Phys.}]
			\step[fieldsource=journal, match={Annual Review of Nuclear and Particle Science},
				replace={Annu. Rev. Nucl. Part. S.}]
			\step[fieldsource=journal, match={The European Physical Journal C},
				replace={Eur. Phys. J. C}]
			\step[fieldsource=journal, match={The European Physical Journal A},
				replace={Eur. Phys. J. A}]
			\step[fieldsource=journal, match={Physics Reports},
				replace={Phys. Rep.}]
			\step[fieldsource=journal, match={Annual Review of Nuclear Science},
				replace={Annu. Rev. Nucl. Sci.}]
			\step[fieldsource=journal, match={Physics Procedia},
				replace={Phys. Proc.}]
			\step[fieldsource=journal, match={Nuclear Physics},
				replace={Nucl. Phys.}]
			\step[fieldsource=journal, match={Nuclear Physics A},
				replace={Nucl. Phys. A}]
			\step[fieldsource=journal, match={Nuclear Physics B},
				replace={Nucl. Phys. B}]
			\step[fieldsource=journal, match={Astronomy and Astrophysics},
				replace={Astron. Astrophys.}]
			\step[fieldsource=journal, match={Atomic Data and Nuclear Data Tables},
				replace={Atom. Data Nucl. Data}]
			\step[fieldsource=journal, match={Physica B: Condensed Matter},
				replace={Physica B}]
			\step[fieldsource=journal, match={The Journal of Chemical Physics},
				replace={J. Chem. Phys.}]
			\step[fieldsource=journal, match={Reports on Progress in Physics},
				replace={Rep. Prog. Phys.}]
			\step[fieldsource=journal, match={Applied Radiation and Isotopes},
				replace={Appl. Radiat. Isotopes}]
			\step[fieldsource=journal, match={Chinese Physics C},
				replace={Chinese Phys. C}]
			\step[fieldsource=journal, match={Journal of Applied Physics},
				replace={J. Appl. Phys}]
			\step[fieldsource=journal, match={Radiation Research},
				replace={Radiat. Res}]
		}
	\map{
		\step[fieldsource=pages,null]
      		\step[fieldsource=doi,final]
      		\step[fieldset=url,null] % set URL to null if doi is found
    		}
	\map{
     		\step[fieldsource=issue, match=\regexp{\A[0-9]+\Z}, final]
     		\step[fieldset=number, origfieldval]
     		\step[fieldset=issue, null]
    		}
	}
}

\AtEveryBibitem{%
  \ifentrytype{article}{%
    \clearfield{pages}%
    \clearfield{month}
  }{\clearfield{pages}%
  }%
}

\newcommand{\cevns}{\protect{CE$\nu$NS}\xspace}

\newcommand{\csi}{CsI{[Na]}\xspace}

\newcommand{\etal}{\emph{et al.}\xspace}

\newcommand{\iso}[2]{\protect{\ensuremath{{}^{#1}\textrm{#2}}}\xspace}

\newcommand{\zprime}{\protect{\ensuremath{Z^\prime}}\xspace}

\newcommand{\nue}{\protect{\ensuremath{\nu_e}}\xspace}

\makeatletter
\newcommand{\printfullauthors}[0]{
	\@author
}
\makeatother

\newcommand\chapterauthor[1]{\authortoc{#1}\printchapterauthor{#1}}
\WithSuffix\newcommand\chapterauthor*[1]{\printchapterauthor{#1}}

\makeatletter
\newcommand{\printchapterauthor}[1]{%
  {\parindent0pt\vspace*{-25pt}%
  \linespread{1.1}\large\scshape#1%
  \par\nobreak\vspace*{35pt}}
  \@afterheading%
}
\newcommand{\authortoc}[1]{%
  \addtocontents{toc}{\vskip-10pt}%
  \addtocontents{toc}{%
    \protect\contentsline{chapter}%
    {\hskip1.3em\mdseries\scshape\protect\scriptsize#1}{}{}}
  \addtocontents{toc}{\vskip5pt}%
}
\makeatother

\newcommand\chaptercoauthor[1]{\printchaptercoauthor{#1}}
\WithSuffix\newcommand\chaptercoauthor*[1]{\printchaptercoauthor{#1}}

\makeatletter
\newcommand{\printchaptercoauthor}[1]{%
  {\parindent0pt\vspace*{-25pt}%
  \linespread{1.1}Co-author(s):~\scshape#1%
  \par\nobreak\vspace*{35pt}}
  \@afterheading%
}
\makeatother

\newcommand\chaptercollab[1]{\printchaptercollab{#1}}
\WithSuffix\newcommand\chaptercollab*[1]{\printchaptercollab{#1}}

\makeatletter
\newcommand{\printchaptercollab}[1]{%
  {\parindent0pt\vspace*{-25pt}%
  \linespread{1.1}\slshape For the #1 collaboration%
  \par\nobreak\vspace*{35pt}}
  \@afterheading%
}
\makeatother

\newcommand\chapterdoi[1]{\doitoc{#1}\printchapterdoi{#1}}
\WithSuffix\newcommand\chapterdoi*[1]{\printchapterdoi{#1}}

\makeatletter
\newcommand{\printchapterdoi}[1]{%
  {\parindent0pt\vspace*{-25pt}%
  \linespread{1.1}\small\scshape Presentation and citeable DOI: \href{http://dx.doi.org/#1}{\protect{\texttt{#1}}}%
  \par\nobreak\vspace*{35pt}}
  \@afterheading%
}
\newcommand{\doitoc}[1]{%
  \addtocontents{toc}{\vskip-15pt}%
  \addtocontents{toc}{%
    \protect\contentsline{chapter}%
    {\hskip1.3em\href{http://dx.doi.org/#1}{\protect{\mdseries\scshape\protect\scriptsize\texttt{#1}}}}{}{}}
  \addtocontents{toc}{\vskip5pt}%
}
\makeatother
\newcommand\chapteraffil[1]{\printchapteraffil{#1}}

\makeatletter
\newcommand{\printchapteraffil}[1]{%
  {\parindent0pt\vspace*{-25pt}%
  \linespread{1.1}\small\slshape #1%
  \par\nobreak\vspace*{35pt}}
  \@afterheading%
}
\makeatother

\titleformat{\chapter}[display]
  {\normalfont\bfseries}{}{0pt}{\Large}

\author[1,2]{D. Aristizabal Sierra}
\author[3]{A.B. Balantekin}
\author[4]{D. Caratelli}
\author[5]{B. Cogswell}
\author[6,7]{J.I. Collar}
\author[8]{C.E. Dahl}
\author[9]{J. Dent}
\author[10]{B. Dutta}
\author[11]{J. Engel}
\author[4]{J. Estrada}
\author[12]{J. Formaggio}
\author[13]{S. Gariazzo}
\author[14]{R. Han}
\author[15,16]{S. Hedges}
\author[17]{P. Huber}
\author[18,19]{A. Konovalov}
\author[20]{R.F. Lang}
\author[10]{S. Liao}
\author[21]{M. Lindner}
\author[4]{P. Machado}
\author[10]{R. Mahapatra}
\author[22]{D. Marfatia}
\author[4,8]{I. Martinez-Soler}
\author[23]{O. Miranda}
\author[24]{D. Misiak}
\author[25]{D.V. Naumov}
\author[26]{J. Newby}
\author[27]{J. Newstead}
\author[13]{D. Papoulias}
\author[28]{K. Patton}
\author[29]{S. Pereverzev}
\author[30,31]{M. Pospelov}
\author[15]{K. Scholberg}
\author[15]{G. Sinev}
\author[32]{R. Strauss}
\author[10]{L. Strigari}
\author[33]{R. Tayloe}
\author[4]{J. Tiffenberg}
\author[34]{M. Vidal}
\author[35]{M. Vignati}
\author[36]{V. Wagner}
\author[9]{J. Walker}
\author[37]{T.-T. Yu}
\author[33]{J. Zettlemoyer}
\affil[1]{Universidad T\'ecnica Federico Santa Mar\'{i}a-Departamento de F\'{i}sica Casilla 110-V, Avda. Espa\~na 1680, Valpara\'{i}so, Chile}
\affil[2]{IFPA, D\'epartment AGO, Universit\'e de Li\`ege, B\^at B5, Sart Tilman B-4000 Li\`ege 1, Belgium}
\affil[3]{Department of Physics, University of Wisconsin, Madison, WI 53706, USA}
\affil[4]{Fermi National Accelerator Laboratory, Batavia, IL 60510, USA}
\affil[5]{The University of Manchester, Manchester, M13 9PL, UK}
\affil[6]{Enrico Fermi Institute and Kavli Institute for Cosmological Physics, University of Chicago, Chicago, IL 60637, USA}
\affil[7]{Department of Physics, University of Chicago, Chicago, IL 60637, USA}
\affil[8]{Northwestern University, Evanston, IL 60208, USA}
\affil[9]{Department of Physics, Sam Houston State University, Huntsville, TX 77341, USA}
\affil[10]{Mitchell Institute for Fundamental Physics and Astronomy, Department of Physics and Astronomy, Texas A\&M University, College Station, TX 77845, USA}
\affil[11]{Department of Physics and Astronomy, University of North Carolina at Chapel Hill, Chapel Hill, NC 27599, USA}
\affil[12]{Massachusetts Institute of Technology, Cambridge, MA 02139, USA}
\affil[13]{Instituto de F\'isica Corpuscular (CSIC-Universitat de Val\`encia), Paterna (Valencia), Spain}
\affil[14]{Beijing Institute of Spacecraft Environment Engineering, Beijing 100094, China}
\affil[15]{Duke University, Durham, NC 27708, USA}
\affil[16]{Triangle Universities Nuclear Laboratory, Durham, NC 27708, USA}
\affil[17]{Center for Neutrino Physics, Virginia Tech, Blacksburg, VA 24061, USA}
\affil[18]{Institute for Theoretical and Experimental Physics named by A.I. Alikhanov of National Research Centre ``Kurchatov Institute'', Moscow, 117218, Russian Federation}
\affil[19]{National Research Nuclear University MEPhI (Moscow Engineering Physics Institute), Moscow, 115409, Russian Federation}
\affil[20]{Purdue University, West Lafayette, IN 47907, USA}
\affil[21]{Max-Planck-Institut f\"ur Kernphysik, Postfach 103980, D-69029 Heidelberg, Germany}
\affil[22]{Department of Physics and Astronomy, University of Hawaii at Manoa, Honolulu, HI 96822, USA}
\affil[23]{Departamento de F\'isica, Centro de Investigaci\'on y de Estudios Avanzados del IPN, Apdo. Postal 14-740, 07000 Ciudad de M\'exico, M\'exico}
\affil[24]{Univ Lyon, Universit\'e Lyon 1, CNRS/IN2P3, IPNL-Lyon, F-69622 Villeurbanne, France}
\affil[25]{Joint Institute for Nuclear Research, Dubna, Russian Federation}
\affil[26]{Oak Ridge National Laboratory, Oak Ridge, TN 37831, USA}
\affil[27]{Department of Physics, Arizona State University, Tempe, AZ 85287, USA}
\affil[28]{Institute for Nuclear Theory, University of Washington, Seattle, WA 98195, USA}
\affil[29]{Lawrence Livermore National Laboratory, Livermore, CA 94550, USA}
\affil[30]{Perimeter Institute for Theoretical Physics, Waterloo, ON, N2J 2W9, Canada}
\affil[31]{Department of Physics and Astronomy, University of Victoria, Victoria, BC, V8P 1A1, Canada}
\affil[32]{Technical University of Munich, D-85748 Garching, Germany}
\affil[33]{Indiana University, Bloomington, IN 47405, USA}
\affil[34]{Department of Physics, Engineering Physics \& Astronomy, Queen's University, Kingston, Ontario K7L 3N6, Canada}
\affil[35]{INFN, Sezione di Roma, P.le Aldo Moro 2, 00185 Roma, Italy}
\affil[36]{IRFU, CEA, Universit\'{e} Paris Saclay, F-91191 Gif-sur-Yvette, France}
\affil[37]{University of Oregon, Eugene, OR 97403, USA}

\begin{document}

% FOR MAKING BACKGROUND IMAGE
% FROM https://tex.stackexchange.com/questions/46280/how-to-create-a-background-image-on-title-page-with-latex
\AddToShipoutPicture*{\BackgroundPic}

\frontmatter

\begin{titlepage}
	\thispagestyle{empty}
	\clearpage
	\newgeometry{left=1in, top=1in, right=1in, bottom=1in}
	
	\begin{center}
	
	%\vspace*{9em}
	{\LARGE Proceedings of\par}
	\vspace{\baselineskip}
	{\Huge The Magnificent \cevns Workshop 2018\par}
	%\vspace{2\baselineskip}

	%\vspace{0.4\paperheight}
	%With contributions from\par
	%\printfullauthors\par
	%\vfill
	\vspace{0.5\paperheight}
	{\large {A workshop held at the University of Chicago\par November 2 \& 3, 2018}\par}
	\vspace{\baselineskip}
	{\Large Editors\par
	G.C. Rich, L. Strigari}
	\end{center}
	
	\pagebreak
	\thispagestyle{empty}
	\clearpage
	\begin{center}
	\textit{With contributions from}\par
	\printfullauthors\par
	\vfill
	\end{center}
\end{titlepage}

\chapter*{Foreword}

\section*{Origins and spirit of the meeting}

From its initial description, the process of coherent elastic neutrino-nucleus scattering (\cevns) has been of interest to the nuclear, particle, and astrophysics communities, appealing to those working either on theory or experiment.
Seizing on some of the momentum generated by the first observation of \cevns by COHERENT in 2017, the Magnificent \cevns workshop was held on November 2 \& 3, 2018 on the campus of the University of Chicago, organized by G.C. Rich, L. Strigari, and J.I. Collar with financial and administrative support generously contributed by the {\bfseries Enrico Fermi Institute (EFI)} and the {\bfseries Kavli Institute for Cosmological Physics (KICP)}, both at the University of Chicago.
Recognition of the interest for such a meeting grew out of conversations at various conferences and workshops in 2018, including: \href{https://www.mpi-hd.mpg.de/nu2018/}{Neutrino 2018} in Heidelberg, Germany; \href{https://indico.ibs.re.kr/event/212/}{NDM 2018} in Daejeon, South Korea; and \href{http://www.ichep2018.org/}{ICHEP 2018} in Seoul, South Korea.
We are grateful to the organizers of the aforementioned meetings for having assembled excellent communities to foster productive discussions, and we would like to especially acknowledge discussions with, and encouragement from, Raimund Strauss.

The goal of this workshop was to develop and strengthen connections between experimentalists and theorists/phenomenologists working in this nascent and exciting field.
By forming strong lines of communications between the many groups working on or around the process, it was hoped that efforts ranging from short- to long-term time scales could be positively impacted, maximizing the realizable scientific output of the community: experimentalists could share with each other lessons learned, helping advance the ongoing experimental efforts; the experimental community could share with theorists/phenomenologists the needed information to most meaningfully incorporate the experimental projects in theoretical calculations; and the theory/pheno community could provide input for next-generation experiments, making sure they target the most exciting physics questions.
Complementing the direct scientific impacts, by encouraging constructive discourse and interaction, we hope to foster an overall positive and inclusive atmosphere in the \cevns community.

This collection of brief summaries\footnote{Credit for the title-page image goes to Connor Awe of Duke University.}, and the accompanying presentations, are meant to serve as a snapshot of the \cevns field as of late 2018.
It is hoped that the Zenodo community for the workshop\footnote{The organizers of Neutrino 2018 set an excellent example for how ``proceedings'' of conferences or workshops could be defined moving forward, making use of the tremendous resource that is Zenodo; see their community at \protect{\href{https://zenodo.org/communities/neutrino2018}{\protect{\texttt{https://zenodo.org/communities/neutrino2018}}}}.}, collecting this document and the presentations, provides a convenient resource for those interested in the process and seeking either high-level or low-level details on the progress of the myriad efforts.

\section*{Citing these proceedings}

In general, the most appropriate way to cite the scientific content of any specific contribution in these proceedings is to {\bfseries reference the Zenodo posting of the contribution in question and to use its particular DOI}.
If making reference to specific scientific content of any contribution, this collection of summaries can be cited {\slshape in addition to contribution-specific Zenodo post} but this decision is left to the discretion of researchers making the reference.

In addition to potentially supplementing a reference to specific contributions, there may be instances where reference to this document by itself are appropriate.
In any case, after replacing {\color{red}\tt{xxxxx}} with the appropriate arXiv identifier, we recommend citing this document as 
\begin{displayquote}
	D. Aristizabal Sierra \etal (2019). Proceedings of The Magnificent \cevns Workshop 2018. Zenodo. DOI: 10.5281/zenodo.3489190. arXiv: {\tt 1910.{\color{red}xxxxx} [hep-ex]}.
\end{displayquote}

\section*{Magnificent \cevns 2019 (and beyond)}

With the success of this meeting, a second workshop on the same subject will be held \textbf{November 9 -- 11, 2019 in Chapel Hill, NC}.
More information can be found at the website \href{http://magnificentcevns.org/2019}{\texttt{magnificentcevns.org/2019}}.
The goal is to make this workshop a regular venue for community building and collaboration between researchers involved, either directly or indirectly, with this exciting and rapidly developing field.

Looking forward, we hope this meeting is able to continue to help direct global \cevns efforts towards the richest scientific program possible and to play a role in facilitating the extraction of exciting new insights from the vibrant palette of experimental results that are expected soon.
Meetings such as this, and strong communities such as that working on \cevns, present opportunities to enhance synergistic activities and align disparate efforts from networks of researchers around the world.
We hope the spirit of collaboration continues to thrive within the \cevns community and involved researchers work to establish best practices / principals to further enhance communication and the sharing of both theoretical and experimental results, along with effective analysis practices honed by the diverse experiences of the community.

\cevns was unobserved for more than 40 years after its description, and efforts to observe the process were originally characterized as ``act[s] of hubris'' \cite{freedman74}.
These experiments remain extremely challenging, and even the act of sharing experimental results must be given thoughtful effort for maximal efficacy \cite{chen2019open}, but the potential in this field to reach for new physics, or to approach questions from different angles and will different probes, is thrilling.
We are optimistic and excited about the science that we expect to be produced by the \cevns community and similarly optimistic that the community dynamic will continue to be welcoming and collaborative, hopefully in even more concrete ways.

\tableofcontents

\mainmatter

% Make the copyright page
%\makecopyrightpage

\chapter{Discrepancies in the published expressions for the \cevns cross section}
\chapterauthor{Alexey Konovalov}

\chapteraffil{Institute for Theoretical and Experimental Physics named by A.I. Alikhanov of National Research Centre ``Kurchatov Institute'', Moscow, 117218, Russian Federation}
\chapteraffil{National Research Nuclear University MEPhI (Moscow Engineering Physics Institute), Moscow, 115409, Russian Federation}

\chapterdoi{10.5281/zenodo.3462599}

Following the first observations the investigation of \cevns is to enter the era of precision measurements. A wide spectrum of scientific goals including detailed study of the nuclear neutron form-factor, non-standard interactions of neutrino with quarks and the electroweak mixing angle value at the energy scale of tens of MeV requires percent and sub-percent precision from both experimental results and the Standard model prediction. The published SM \cevns cross section predictions including the recent ones lack the comprehensive expression taking into account effects of the weak axial current and the spin of a nucleus. It is of particular interest if the transitions with a change of the nuclear spin projection should contribute to a coherent or an inelastic channel. A comprehensive expression for the SM \cevns cross section and corresponding calculation apparatus are highly desirable in order to encourage the experimental effort and improve understanding of various apects of \cevns.
\chapter{Revisiting the axial contribution to \cevns}
\chapterauthor{Jayden Newstead}

\chapteraffil{Department of Physics, Arizona State University, Tempe, AZ 85287, USA}

\chapterdoi{10.5281/zenodo.3462607}

A precise calculation of the Standard Model \cevns rate is required before any discovery of `new physics' can be claimed. With the increasing number of experimental groups entering the \cevns community, bringing diverse detector targets and neutrino sources, it is desirable to have a consistent formalism for predicting experimental rates.  As pointed out in A.~Konovalov's talk, there are discrepancies among \cevns cross sections in the literature. Additionally, there are few examples in the literature which account for the axial currents. 
The appropriate formalism for these calculations in semi-leptonic electroweak theory was originally developed in \cite{DeForest:1966ycn} and \cite{Donnelly:1979}. In this formalism the hadronic currents are spherically decomposed and expanded in multipoles to obtain irreducible tensor operators which act on single nucleon states, which can be expressed in a harmonic oscillator basis~\cite{Donnelly:1979ezn}. Given the low momentum transfer of the \cevns process, the one-body calculation provides a reasonable starting point (and can be efficiently calculated using available tools). As an example I have calculated the rate expected by the COHERENT collaboration, finding that the axial contribution is negligible. Previous calculations overestimate the axial contribution for two reasons, first they do not properly handle the projection of the neutrino's spin onto the nuclear spin, and second, the spin held by the nucleons is overestimated.

\chapter{Neutrino Scattering to Understand ``$\mathbf{g_A}$ Quenching''}
\chapterauthor{Jon Engel}

\chapteraffil{Department of Physics and Astronomy, University of North Carolina at Chapel Hill, Chapel Hill, NC 27599, USA}

\chapterdoi{10.5281/zenodo.3462666}

Calculations within nuclear models overestimate the matrix elements that occur
in beta decay, two-neutrino double-beta decay, and related processes.  The
overestimate is often remedied by using an artificially reduced (``quenched'')
value for the axial-vector coupling constant $g_A$, which multiplies the matrix
elements.  The physics responsible for the quenching is not fully understood,
but must be a combination of correlations that escape models and many-body weak
currents.  The momentum dependence of the quenching, which we need to know to
accurately calculate the matrix elements that govern neutrinoless double-beta
decay, will depend on the relative size of these two contributions.  

Neutrinos from stopped pions can transfer significant amounts of momentum.
Measuring the cross sections for inelastic neutrino-nucleus scattering thus has
the potential to tell us about the momentum dependence of $g_A$ quenching.  In
charge-current scattering, one can obtain detailed information on inelastic
events by measuring the energies of outgoing electrons.  If the energies of any
photons can be measured as well, then one can reconstruct cross sections to
specific excited states and compare them with the predictions of models and/or
ab-initio calculations.  The data would be of immense help to theorists
struggling to reduce the currently large uncertainty in double-beta matrix
elements.

\chapter{Coherency and incoherency in neutrino-nucleus elastic and inelastic scattering}

\chapterauthor{Dmitry V. Naumov}

\chapteraffil{Joint Institute for Nuclear Research, Dubna, Russian Federation}

\chaptercoauthor{Vadim A. Bednyakov}

\chapterdoi{10.5281/zenodo.3464596}

Neutrino-nucleus scattering $\nu A\to \nu A$, in which the nucleus conserves its integrity, is considered.
Our consideration follows a microscopic description of the nucleus as a bound state of its constituent nucleons described by a multi-particle wave-function of a general form.

We  show  that elastic interactions keeping the nucleus in the same quantum state lead to a quadratic enhancement of the corresponding cross-section in terms of the number of nucleons.
Meanwhile, the cross-section of inelastic processes in which the quantum state of the nucleus is changed, essentially has a linear dependence on the number of nucleons.
These two classes of processes are referred to as coherent and incoherent, respectively.

Accounting for all possible  initial and final internal states of the nucleus leads to a general conclusion independent of the nuclear model.
The coherent and incoherent cross-sections are driven by factors $|F_{p/n}|^2$ and $(1-|F_{p/n}|^2)$, where $|F_{p/n}|^2$ is a proton/neutron form-factor of the nucleus, averaged over its initial states.
Therefore, our assessment suggests a smooth transition between regimes of coherent and incoherent neutrino-nucleus scattering.
In general, both  regimes contribute to experimental observables.

The coherent cross-section formula used in the literature is revised and corrections depending on kinematics are estimated.
Consideration of only those matrix elements  which correspond to the same initial and final  spin states of the nucleus and accounting for a non-zero momentum of the target nucleon are two main sources of the corrections.

As an illustration of the importance of the incoherent channel we considered three experimental setups with different nuclei.
As an  example, for ${}^{133}\text{Cs}$ and neutrino energies of $30-50$ MeV the incoherent cross-section is about 10-20\% of the coherent contribution if experimental detection threshold is accounted for.

Experiments attempting to measure coherent neutrino scattering by solely detecting the recoiling nucleus, as is typical, might be including an incoherent background that is indistinguishable from the signal if the excitation gamma eludes its detection. 
However, as is shown here, the incoherent component can be measured directly by searching for photons released by the excited nuclei inherent to the incoherent channel.
For a beam experiment these gammas should be correlated in time with the beam, and their higher energies make the corresponding signal easily detectable at a rate governed by the ratio of incoherent to coherent cross-sections.
The detection of signals due to the nuclear recoil and excitation $\gamma$s provides a more sensitive instrument in studies of nuclear structure and possible signs of new physics.
\chapter{Constraining NSI with Multiple Targets}
\chapterauthor{Gleb Sinev}

\chapteraffil{Duke University, Durham, NC 27708, USA}

\chapterdoi{10.5281/zenodo.3464645}

\begin{figure}[h]
\centering \includegraphics[bb= 0 0 461 346, width=0.75\textwidth]{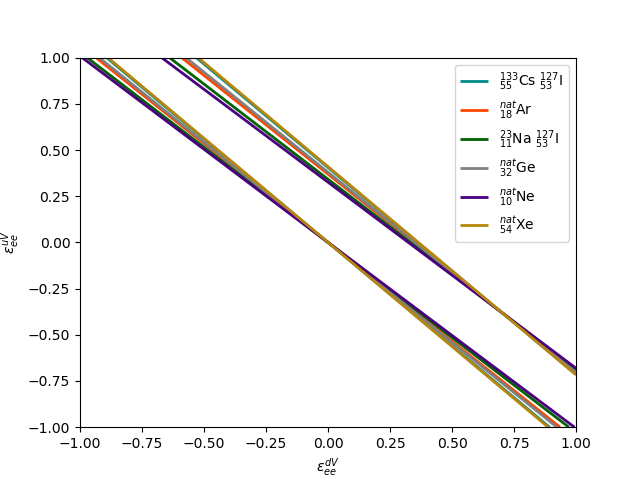}
\caption{Values of two NSI couplings (with the rest of the couplings set to 0)
  that do not change the Standard Model \cevns rate
  for each of the considered target materials.}
\label{fig:nsi_modification}
\end{figure}

Non-standard interactions (NSI) mediated by heavy particles can suppress
or enhance the rate of coherent elastic neutrino-nucleus scattering (\cevns)
by introducing additional couplings between neutrinos
and quarks~\cite{Barranco:2005}.
The values of these couplings can be constrained by \cevns measurements;
however, using a single target nucleus for this purpose results in
ambiguities, because different combinations of coupling values
can produce the same detected rate of nuclear recoils.
Therefore, \cevns detection on multiple targets is required to make
an unambiguous measurement of the NSI couplings, with a combination
of light and heavy nuclei providing the best result.
This contribution shows, as an example, NSI coupling values producing
Standard Model \cevns rates in six targets:
CsI, Ar, NaI, Ge, Ne, and Xe (see figure~\ref{fig:nsi_modification}).
The COHERENT experiment has been taking \cevns data with CsI and Ar detectors
and has plans to install NaI and Ge detectors
in the near future~\cite{Akimov:2018},
a combined analysis of which can result in a significant decrease of
the allowed parameter space for the NSI couplings.

% fig bounding box
% 0 0 461 346
% 

\chapter{Constraints on neutrino generalized interactions from COHERENT data}
\chapterauthor{Diego {Aristizabal Sierra}}

\chapteraffil{Universidad T\'ecnica Federico Santa Mar\'{i}a-Departamento de F\'{i}sica Casilla 110-V, Avda. Espa\~na 1680, Valpara\'{i}so, Chile}
\chapteraffil{IFPA, D\'epartment AGO, Universit\'e de Li\`ege, B\^at B5, Sart Tilman B-4000 Li\`ege 1, Belgium}

\chaptercoauthor{V. De Romeri, N. Rojas}

\chapterdoi{10.5281/zenodo.3463490}

Neutrino Generalized Interactions (NGI) are dimension-six effective neutrino-quark interactions which cover all possible Lorentz invariant structures.
From the effective point of view NGI determine the most general way of accessing new interactions effects in \cevns experiments. 
They include the well-studied neutrino non-standard interactions (NSI), but span a larger set which includes --- among others --- scalar and tensor four-point couplings.
NGI are constrained by data from laboratory experiments which include neutrino deep inelastic scattering (CHARM and NuTeV) and monojets searches at the LHC.
Constraints from the former are readily evaded if the NGI are generated by mediators with masses below 1 GeV or so, while limits from the latter are relevant if the mediator can be integrated out at LHC energies, $\mathcal{O}(E)\sim \text{TeV}$.
Thus if one relies only on these laboratories probes the mediator mass range $[1,10^3]\,$MeV is barely unconstrained. 
From that perspective, COHERENT data plays a crucial role. 
Bounds on NGI derived from the observation of the \cevns process place sensitive bounds on the otherwise poorly constrained parameter space.
The limits, although substantial, still enable for rather large NGI parameters, this mainly due to the relatively large experimental uncertainties. 
Future COHERENT measurements (or any other \cevns measurement, with uncertainties below those currently involved) can either observe effects of these new interactions or improve on NGI limits.
Near-future experimental setups such as COHERENT phase-II and phase-III, CONUS and multi-tonne scale dark matter detectors (e.g. DARWIN or DarkSide-20k) will be able to test whether traces of NGI are present in \cevns.

{\slshape [Editor's note: the work summarized here and reflected in the associated presentation is published as Ref. \cite{AristizabalSierra:2018eqm}.]}
\chapter{NSI @ \cevns etc}
\chapterauthor{Danny Marfatia}

\chapteraffil{Department of Physics and Astronomy, University of Hawaii at Manoa, Honolulu, HI 96822, USA}

\chapterdoi{10.5281/zenodo.3464543}

Coherent elastic neutrino-nucleus scattering consistent with the standard model has been observed by the COHERENT experiment. For nonstandard neutrino interactions (NSI) generated by a vector mediator lighter than 50 MeV, only couplings of the mediator are constrained by the detected spectrum, and large NSI are still viable. For a heavier mediator, in spite of degeneracies between the NSI parameters, current COHERENT data place meaningful constraints on the effective NSI parameters in Earth matter.
\chapter{Model building and connections to charged current experiments}
\chapterauthor{Bhaskar Dutta}

\chapteraffil{Mitchell Institute for Fundamental Physics and Astronomy, Department of Physics and Astronomy, Texas A\&M University, College Station, TX 77845, USA}

\chapterdoi{10.5281/zenodo.3464465}

The recent detection of \cevns events by the COHERENT experiment using 14.6-kg \csi scintillator detectors at 6.7-$\sigma$ level has opened up a new window into the neutrino interactions in the low energy regime, and along with it provides a new probe into beyond the SM physics. Since \cevns is well predicted in the SM and therefore a measured deviation from it can provide a test of new physics.In this section we highlight and discuss three well-motivated scenarios for new physics which have particles in the MeV-GeV range: (i) kinetic mixing, (ii) hidden sectors, and (iii) scenarios with a $L_\mu$-$L_\tau$ symmetry. We highlight the role of \cevns and low energy neutrino experiments in probing these models. In addition, \cevns experiments can also probe the parameter space of a fourth neutrino with mass Delta $m^2 \sim 1 \text{eV}^2$, which has been hinted at by several neutrino experiments but whose existence is still inconclusive. The inclusion of both timing and energy data provide the best constraint for all these models compared to most of the experimental constraints. In addition \cevns experiments can probe light dark matter models with a choice of optimized cuts ($E_\text{rec} >14~\text{keV}$ and $t < 1.5~\mu s$) which would remove the SM background. The ongoing COHERENT constraint on the dark matter parameter space also is better than most of the existing constraints on light dark matter models.
\chapter{Astrophysical Applications of Coherent Neutrino Scattering}
\chapterauthor{Louis Strigari}

\chapteraffil{Mitchell Institute for Fundamental Physics and Astronomy, Department of Physics and Astronomy, Texas A\&M University, College Station, TX 77845, USA}

\chapterdoi{10.5281/zenodo.3462375}

With the recent discovery of coherent elastic neutrino-nucleus scattering, we are now in the exciting position of looking forward to understand the physics that can be extracted from the CE$\nu$NS signal. The CE$\nu$NS process is unique, because it is intertwined with three major areas of physics: nuclear, high-energy, and astrophysics. From a nuclear physics perspective, CE$\nu$NS will shed light on the neutron form factor and the weak charge distribution in the nucleus. From a high-energy physics perspective, because it is able to probe non-standard neutrino interactions (NSI) separately for up and down quarks, CE$\nu$NS will probe NSI in a regime that is not accessible to standard neutrino oscillation experiments. In astrophysics, CE$\nu$NS will provide a new window into the interior of supernovae, the Sun, and the atmosphere. In addition to these theoretical applications, it will be possible to search for sterile neutrinos, and shed light on the reactor and gallium anomalies, which both remain unexplained. 

There are now several experimental probes of CE$\nu$NS that aim to measure this cross section across a wide range of energy scales. These include terrestrial experiments that utilize nuclear reactor and stopped pion sources, as well as astroparticle experiments that search for dark matter. A CE$\nu$NS detection through each of these methods will ultimately be important to understand the energy dependance of the signal, as well as how the signal depends on the target used for detection. Thus there is a natural three-pronged, complementary experimental approach that may be utilized in order to extract maximal information from the signal. 

In the particular field of Solar neutrinos, with a CE$\nu$NS detection dark matter experiments will be able to probe the properties of neutrinos and the Sun that has not been possible with current experiments. For example, future liquid noble gas and cryogenic dark matter experiments will be able to make the first neutral current measurement of the \iso{8}{B} components of the solar neutrino flux. This will provide the first detected neutral current energy spectrum of \iso{8}{B} neutrinos, and be able to shed light on the long-standing Solar metallicity issue. This detection will also help in a search for sterile neutrinos. With low threshold dark matter detectors, lower energy components of the Solar neutrino flux may also be studied. These can be the first pure neutral current measurements of the low-energy \iso{7}{Be} and {\it pep} solar neutrino fluxes. Low-threshold ton-scale detectors may also be able to establish the first measurement of neutrinos from the CNO cycle. This is a long sought-after component of the solar neutrino spectrum that generates $\sim 1$\% of solar energy.
\chapter{Coherent scattering of ``light objects'' on nuclei}
\chapterauthor{Maxim Pospelov}

\chapteraffil{Perimeter Institute for Theoretical Physics, Waterloo, ON, N2J 2W9, Canada}
\chapteraffil{Department of Physics and Astronomy, University of Victoria, Victoria, BC, V8P 1A1, Canada}

\chapterdoi{10.5281/zenodo.3464632}

This talk is based on two recent papers, Refs. \cite{Cui:2017ytb,Bringmann:2018cvk}\footnote{Editor's note: the citation for \cite{Bringmann:2018cvk} was updated, subsequent to the submission of these proceedings, to represent the published work; the original citation referred to the (first) arXiv submission alone.}, and the unifying theme for both is the ``coherent scattering of light obejcts'' on nuclei. In a broad sense, it fits the theme of this meeting, dedicated to the 
neutrino coherent scattering. 

\begin{figure}
\begin{center}
\includegraphics[width=7cm,bb=0 0 270 247]{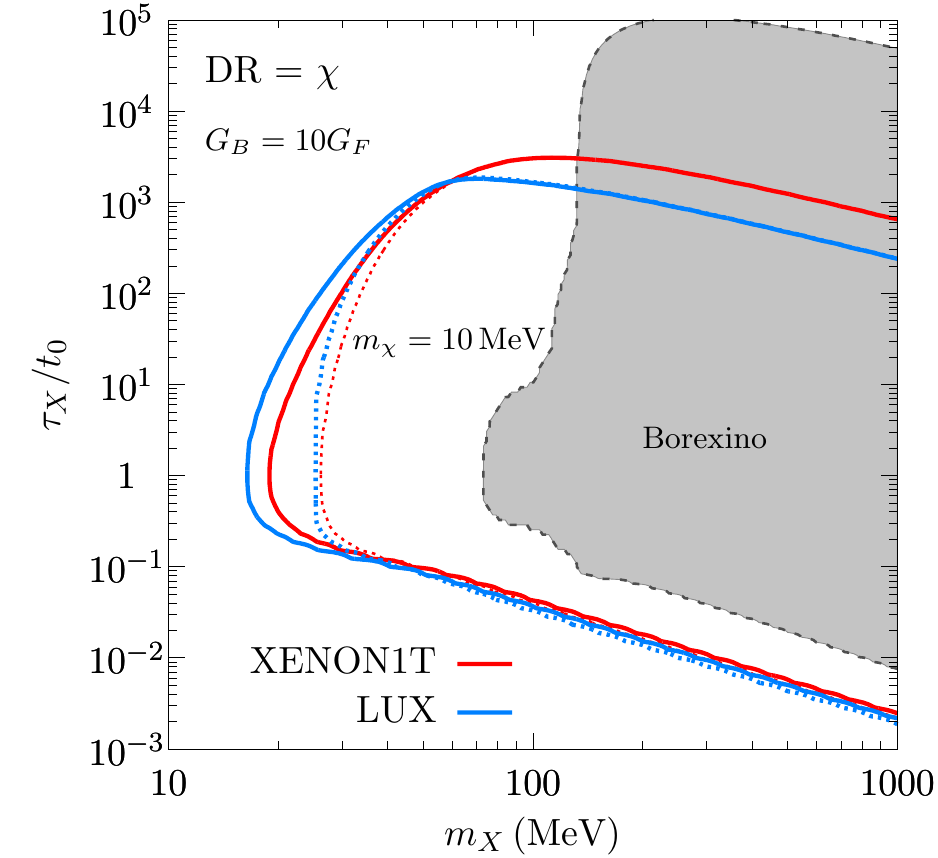}
\end{center}
\caption{\label{pospelov:fig:1}Constraint on the parameter space of decaying dark matter particles $X$. Its products of decay, dark radiation particles $\chi$, 
scatter on nuceli via a baryonic force of strength $G_B$ producing a recoil. }
\end{figure}

The first paper \cite{Cui:2017ytb} considers the case of interacting ``dark radiation'' (DR). Usually dark radiation is mentioned in the framwork 
of modified cosmology, when it contributes to the energy density of the Universe, and modifies the Hubble expansion rate. In our approach, 
we have considered a hypothetical case of dark radiation which constitutes a subdominant fraction of Universe's energy balance, 
and is not numerous, but rather energetic compared to the CMB:
\begin{equation}
\omega_{DR}n_{DR} < \rho_{tot};~~ \omega_{DR} \gg \omega_{CMB}; ~~ n_{DR} \gg n_{CMB}. 
\end{equation}
A concrete realization of such situation occur when massive DM particles decay to dark radiation. The central question studied in our paper is 
about prospects of searching for such dark radiation component, using its coherent scattering on nuclei. It is easy to see that for the DM mass in tens of 
MeV, and the lifetime against decaying to DR somewhat longer than the age of the Universe, the resulting flux of DR particles can be significant and indeed 
comparable to the \iso{8}{B} neutrino flux, $\mathcal{O}(10^6{\rm cm}^{-2}{\rm s}^{-1})$. 
If there is an interaction between DR and nuclei, and DR and electrons, then there is a chance of detecting 
DR prior to detecting dark matter (DM) particles. Our paper considers two types of interaction, via a new light particle called dark photon, and via an analogous vector boson that couples only to baryons. {\em Conclusions}: coherent scattering of DR on nuclei, specifically in the underground ``direct detection'' experiments, provide a very competitive sensitivity reach to DR, and limit its interaction strength with nuclei to be comparable or  smaller than the weak interaction strength, $G_F$. Figure \ref{pospelov:fig:1} summarizes the constraints on the parameter space of the model in case of the baryonic force mediator.

\begin{figure}
\begin{center}
\includegraphics[width=7cm,bb= 0 0 450 304]{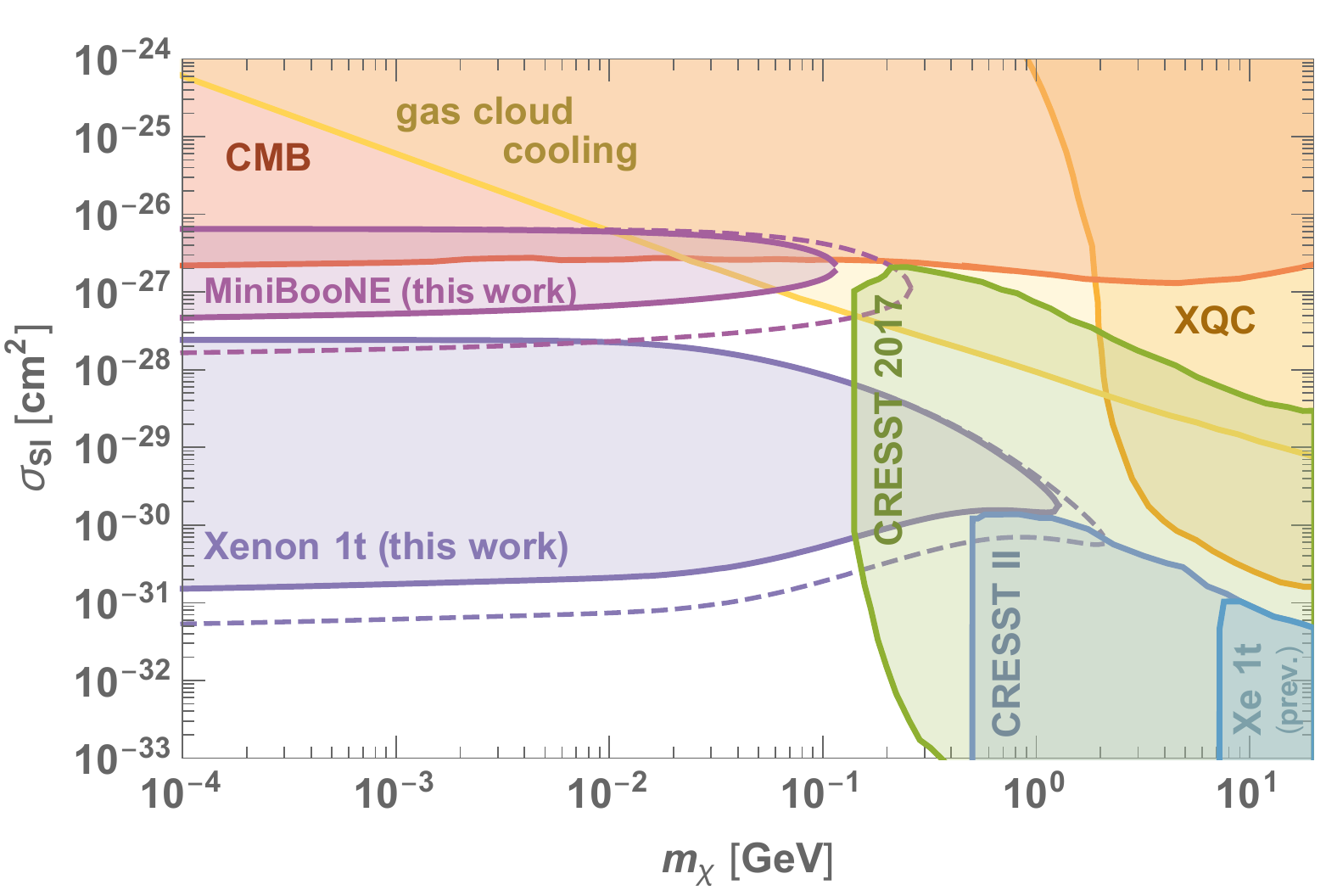}
\end{center}
\caption{\label{pospelov:fig:2}Constraint on the spin-independent scattering cross section derived from the two step-collision: cosmic rays colliding with DM, 
and accelerating it to relativstic velocities, with subsequent scattering of energetic particles within neutrino and DM detectors.}
\end{figure}

Second work presented in this talk \cite{Bringmann:2018cvk}\footnote{Editor's note: see previous.}, is a recent study of light and relatively strongly interacting dark matter.  In this case, 
the signal from dark matter elastic scattering falls below the experimental threshold for detection. Our paper derives novel limits, employing a two-step process. 
First, the cosmic rays collide with dark matter particles, and accelerate them to significant velocities/momenta. These ``cosmic ray dark matter'' (CRDM) states then scatter inside the detectors creating observable signals, as they are safely above the background. When the stopping inside the Earth is not an issue, 
the signal scales as $\propto \sigma_{\chi}^2$. For sizeable $\sigma_\chi$ and small DM masses $m_\chi$, we derive novel limits, 
that reach down to cross sections $10^{-31}$cm$^2$ and apply to all masses, including a very small $m_\chi$. Again, elastic scattering of 
CRDM on nuclei is the main mechanism driving these limits, that for the case of the spin-independent scattering are shown in Figure \ref{pospelov:fig:2}.

\chapter{Sub-GeV Dark Matter Theory}
\chapterauthor{Tien-Tien Yu}

\chapteraffil{University of Oregon, Eugene, OR 97403, USA}

\chapterdoi{10.5281/zenodo.3464705}

In recent years, there has been incredible interest in the search for sub-GeV candidates of particle dark matter (see e.g.~\cite{Battaglieri:2017aum} for summary).
One of the proposed avenues is new techniques for the direct detection of dark matter.
Traditional direct detection experiments which have been in operation for the last few decades are optimized for the detection of weakly-interacting massive particles (WIMPs) that scatter off of the nuclei in the various target materials.
These dark matter candidates are typically ${\cal O}(100{~\rm GeV})$ in mass and result in energy transfers of ${\cal O}(10{~\rm keV})$, which is well within the realm of detectability.
However, once the DM mass drops below a few GeV, the energy transfer drops to below $\sim$ eV and these nuclear recoil experiments completely lose sensitivity.
Instead, for these sub-GeV candidates, one can consider DM-electron scattering, which results in energy transfers of ${\cal O}(10{~\rm eV})$.
The energy transfer manifests itself as ionized electrons or photons, depending on the experimental setup.
Thus, successful experimental setups need to have energy thresholds to detect electrons and photons with energies of around an eV. 

Coherent neutrino-nucleus scattering (\cevns) is an inevitable background to any dark matter direct detection experiment as it closely mimic the signature of dark matter.
For DM-nuclear scattering experiments, the signature of the neutrino-scattering vs. DM-scattering is the same: a recoiling nucleus.
However, \cevns also manifests itself in DM-electron scattering experiments.
The nuclear recoil that results from \cevns will produce secondary electrons, a process whose efficiency can be calculated through various models such as the Lindhard model~\cite{Lindhard}.
For sub-GeV DM experiments, the dominant source of neutrinos are solar neutrinos and the impact of solar neutrinos on sub-GeV DM searches was investigated in~\cite{Essig:2018tss}.
The exposures for detecting at least one neutrino event range from 0.05 to 9.7 kg-years for silicon, germanium, and xenon, with xenon at the low-end of the range and germanium and silicon at the high-end.
The exact value also depends on the ionization efficiency.
For example, the exposures in silicon can range from 0.2-9.7 kg-years, 0.3-2 kg-years for germanium, and 0.05-0.16 kg-years for xenon.
Thus, the sensitivity to a sub-GeV DM search will be limited by neutrinos once the exposures are larger than those listed above.
However, there is no absolute neutrino ``floor'' beyond which there is no improvement possible. 
\chapter{\cevns in dark matter experiments}
\chapterauthor{Pedro Machado}

\chapteraffil{Fermi National Accelerator Laboratory, Batavia, IL 60510, USA}

\chapterdoi{10.5281/zenodo.3464517}

In this talk, I have presented how new physics models could contribute to the coherent elastic neutrino-nucleus scattering cross section. 
New physics could enhance the irreducible solar neutrino background, a.k.a. the neutrino floor, in dark matter experiments. 
By examining the experimentally allowed parameter space in three realistic models, we have estimated the maximum enhancement the neutrino floor could receive. 
The non-standard neutrino floor could easily be a factor of two larger than the standard model case, or even greater depending on the robustness of certain astrophysical constraints.
\chapter{\cevns in the 2020s With Liquid Xenon}
\chapterauthor{Rafael F. Lang}

\chapteraffil{Purdue University, West Lafayette, IN 47907, USA}

\chapterdoi{10.5281/zenodo.3464508}

As searches for WIMP Dark Matter require low ($\sim$keV) energy thresholds, direct Dark Matter detection experiments can also be sensitive to \cevns. Liquid xenon time projection chambers are a particularly successful and promising technology to fully probe the accessible WIMP Dark Matter parameter space. These experiments search for nuclear recoils from simple elastic scatters. Since the relevant kinematics is degenerate in  momentum transferred, these detectors can in principle not distinguish the nuclear recoil spectrum induced by non-relativistic ($10^{-3}$c) heavy ($>$GeV$/c^2$) WIMPs from the spectrum of corresponding light, relativistic neutrinos through \cevns~\cite{Billard:2013qya}. Thus, \cevns from astrophysical neutrino sources is now often shown in the usual WIMP explusion plots. In particular the \cevns signal from atmospheric neutrinos has become known as the \textit{neutrino floor} of direct detection, although the name is rather misleading for a variety of reasons. The currently-running XENON1T experiment~\cite{Aprile:2017aty} is already sensitive to any Galactic supernova through this channel~\cite{Lang:2016zhv}. 2019 will see the commissioning of XENONnT and LZ, which can be expected to measure solar boron-8 neutrinos through \cevns in a few years~\cite{Mount:2017qzi}. Given a low-enough threshold in the usual scintillation-plus-ionization channel, or else a low-enough background in the ionization-only channel as pursued by the LBECA collaboration, these experiments might improve our knowledge of the solar metallicity through this channel~\cite{Robertson:2012ib}. However, truly probing the neutrino floor requires a measurement of \cevns from atmospheric neutrinos. To properly achieve this such a measurement requires an exposure of order 1~kilotonne$\times$year in xenon, only achievable by a generation-3 Dark Matter experiment~\cite{Aalbers:2016jon}. In all cases, measuring \cevns in those experiment has unique sensitivity to a variety of new interactions~\cite{Dutta:2017nht}. Taken together, such liquid xenon experiments will feature a rich science case with signals from various Dark Matter candidates as well as from a variety of astrophysical neutrino sources.
\chapter{Resolving CP degeneracy using atmospheric neutrino at dark matter detector}
\chapterauthor{Shu Liao}

\chapteraffil{Mitchell Institute for Fundamental Physics and Astronomy, Department of Physics and Astronomy, Texas A\&M University, College Station, TX 77845, USA}

\chaptercoauthor{B. Dutta, L. Strigari}

\chapterdoi{10.5281/zenodo.3462628}

Direct dark matter search detectors provide a source to examine the non-standard aspect of neutrino interactions via solar and atmospheric neutrinos.
The low threshold of such detectors will probe the some NSI parameter space at $2\sigma$  significance through solar neutrino with a tonne-year scale exposure. 
It will also allow the observation of the influence of NSI parameters on neutrino oscillation.
Through the oscillation at different zenith angles of atmospheric neutrino, the future observation of atmospheric neutrino at direct dark matter search detector can help to resolve the degeneracies between CP phase and NSI parameters, which is otherwise impossible to solve at fixed length neutrino oscillation experiment.
\chapter{Bremsstrahlung and the Migdal Effect for Coherent Elastic Neutrino-Nucleus Scattering (\protect{\cevns})}
\chapterauthor{James Dent}

\chapteraffil{Department of Physics, Sam Houston State University, Huntsville, TX 77341, USA}

\chaptercoauthor{N. Bell, J. Newstead, S. Sabharwal, T. Weiler}

\chapterdoi{10.5281/zenodo.3464454}

\begin{figure*}[htb]
\centering
\begin{tabular}{cc}
\includegraphics[width=5cm, bb= 0 0 450 478]{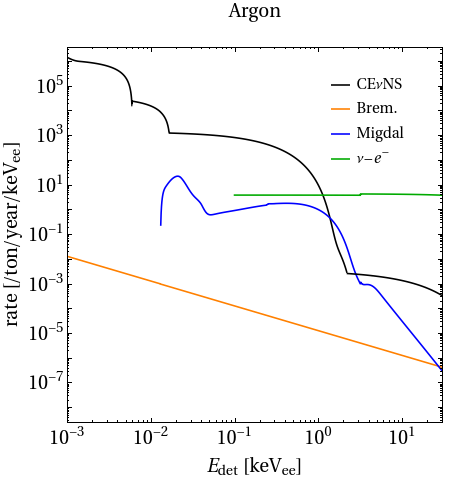} & \includegraphics[width=5cm, bb= 0 0 1000 1062]{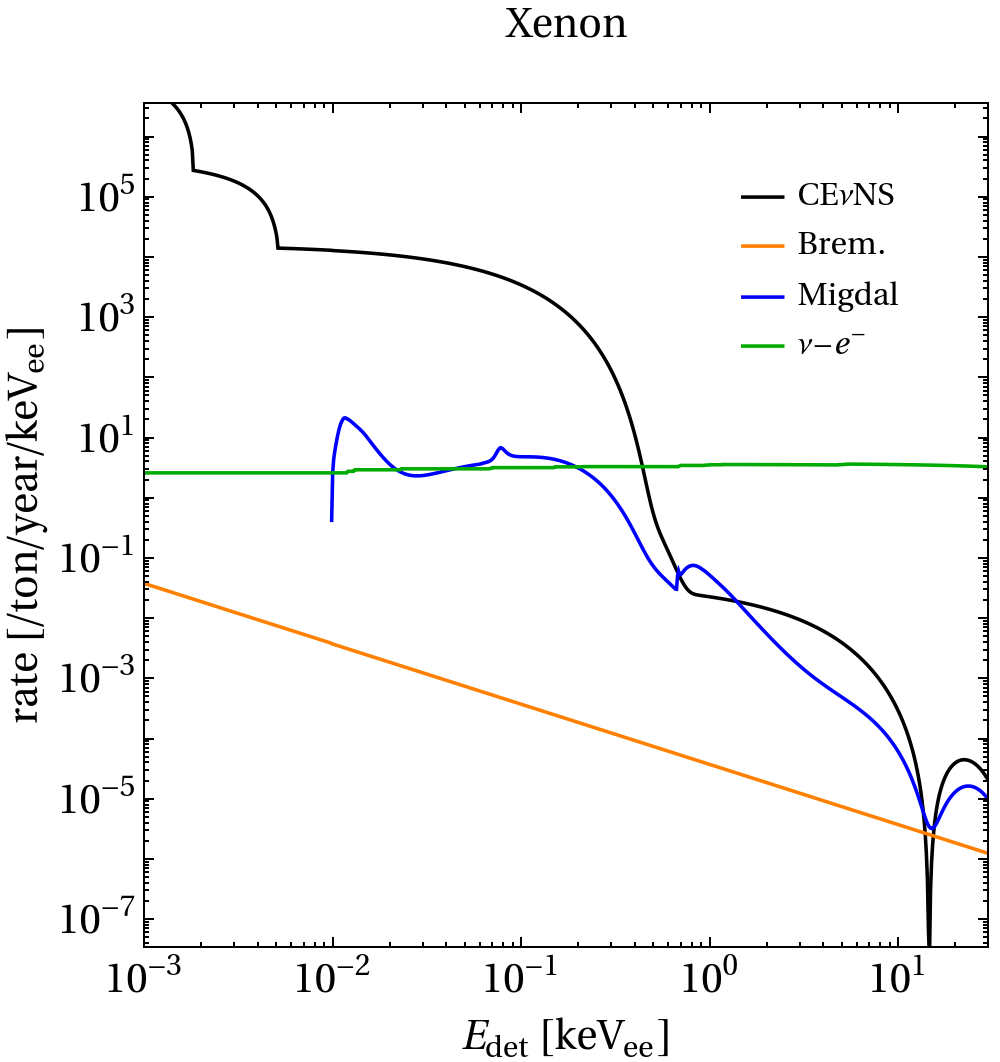} \\
\includegraphics[width=5cm,bb = 0 0 600 628]{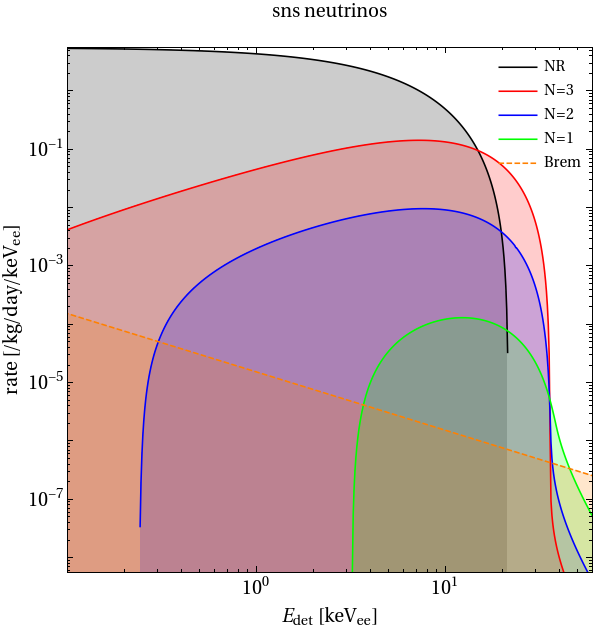} & \includegraphics[width=5cm, bb= 0 0 600 628]{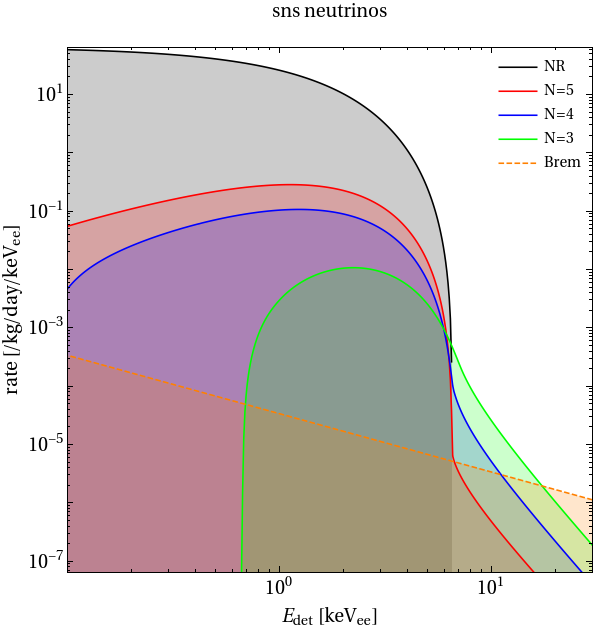} \\
\includegraphics[width=5cm, bb = 0 0 1000 1034]{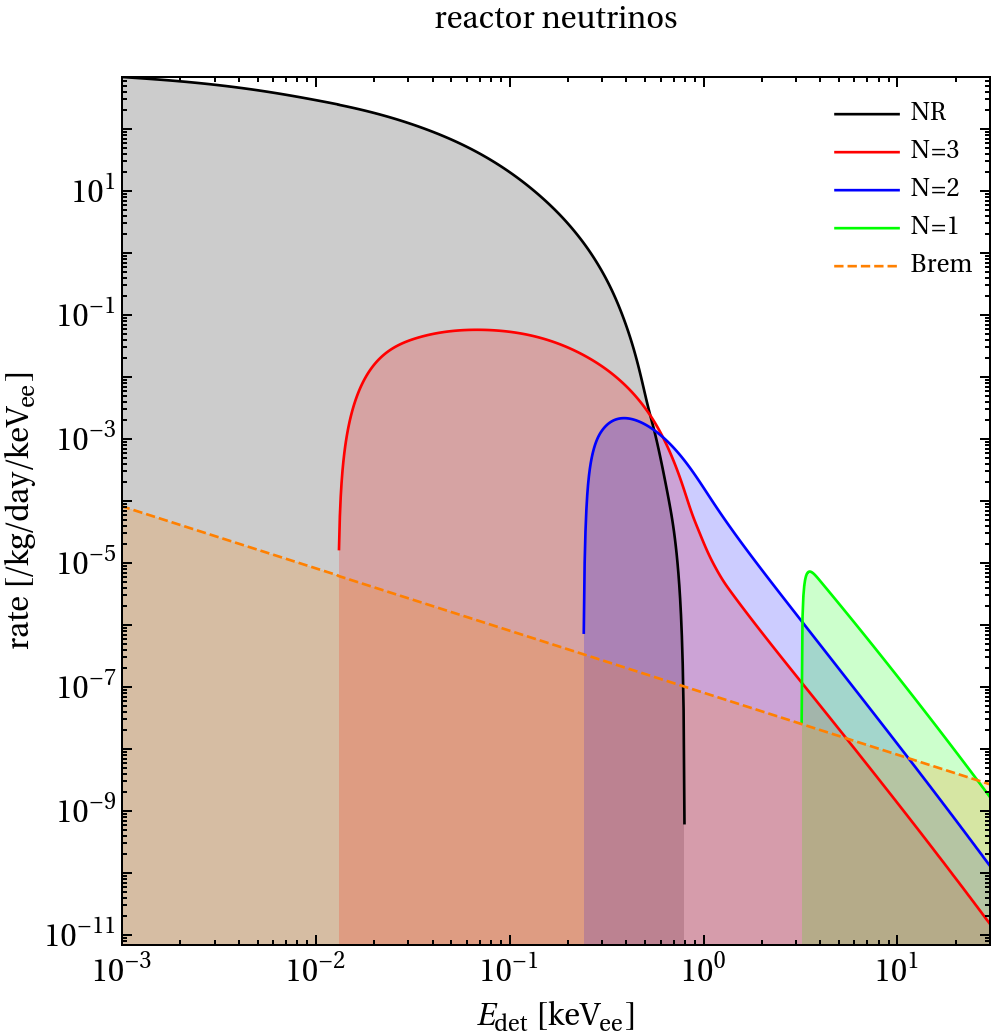} & \includegraphics[width=5cm, bb = 0 0 1000 1046]{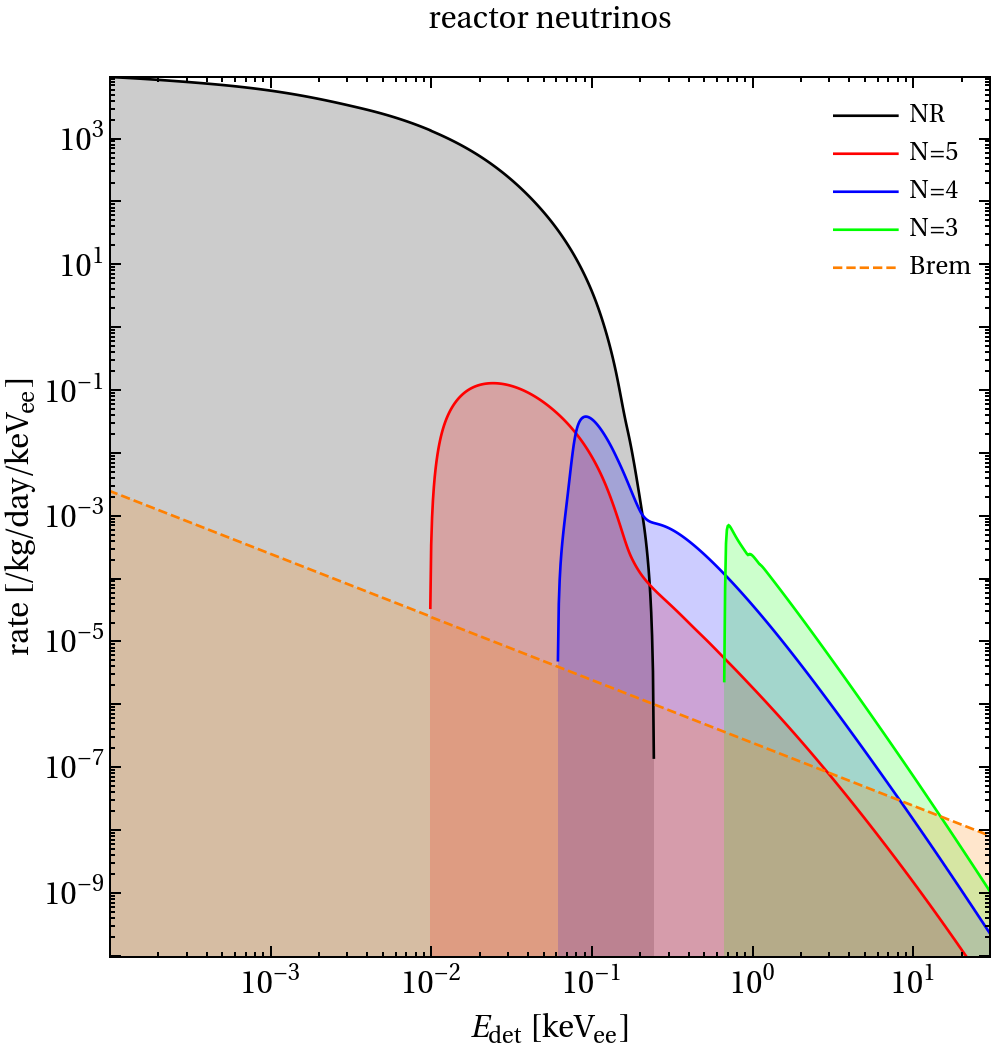} \\
\end{tabular}
\caption{Rates for nuclear recoils (NR), the Migdal effect (where \protect{$N$} denotes the contribution from electrons in the \protect{$N^\text{th}$} energy level), and bremsstrahlung for liquid argon (left) and liquid xenon (right) detectors. The processes addressed are: astrophysical neutrinos (top) with $\nu-e$ scattering also shown, stopped pion sourced neutrinos such as those at the SNS (middle), and reactor neutrinos (bottom) normalized to that of a detector a distance of 1 m from a 1 MW reactor.}
\label{fig:Migdal-brem-rates}
\end{figure*}

It has recently been shown that including the effects of photon bremsstrahlung in the dark matter-nucleus scattering process \cite{Kouvaris:2016afs}, or examining ionization and electronic excitation of a target atom due to the Migdal effect (where the electron cloud's motion is not modeled as instantaneously following the recoiling nucleus \cite{Migdal:1941,Ibe:2017yqa,Akerib:2018hck,Armengaud:2019kfj,Dolan:2017xbu}\footnote{Editor's note: following submission of these proceedings, the reference to \protect{\cite{Dolan:2017xbu}} was updated to represent the published version of the article; the original citation referred only to the arXiv version, which was last revised prior to the Magnificent \protect{\cevns} workshop and, based on revision notes, should match the published work.}\footnote{Editor's note: similarly, Ref. \protect{\cite{Akerib:2018hck}} has been updated to reflect the published work; the reference should be to the initial arXiv version. In this case, it is not clear if there are substantive changes between the initial preprint and the published work.}), can extend the reach of direct detection experiments to lower dark matter masses. It is also well known that, as experiments searching for dark matter through direct detection continue to achieve lower thresholds and larger exposures, they can become sensitive to a solar and atmospheric neutrino background interacting with the detector's target nuclei via the \cevns process. It is therefore of interest to determine whether bremsstrahlung or the Migdal effect accompanying the \cevns process can provide additional experimental signals. In Fig.~\ref{fig:Migdal-brem-rates} we show the effects for the bremsstrahlung process and the Migdal effect in both liquid argon and liquid xenon targets for astrophysical neutrinos (this includes both solar and atmospheric neutrinos), as well as for reactor neutrinos and neutrinos from the stopped pion source at the SNS. We see that the bremsstrahlung process is sub-dominant to both the Migdal effect and the standard nuclear recoil except at energies above $\mathcal{O}\left(10~{\rm{keVee}}\right)$ for reactor and stopped-pion sourced neutrinos, though the rate at those energies is suppressed by many orders of magnitude compared to the peak for nuclear recoils. For astrophysical neutrinos, the Migdal effect has competitive rates with the standard nuclear recoil rate at the point where the \cevns process from atmospheric neutrinos becomes dominant as the \iso{8}{B} neutrino flux rapidly diminishes. This could pose an interesting opportunity for future multi-ton direct detection experiments. 

% bounding boxes
% figure_dent_sun_ar 0 0 450 478
% figure_dent_xe_total 0 0 1000 1062
% figure_dent_sns_ar 0 0 600 628
% figure_dent_sns_xe 0 0 600 628
% figure_dent_reactor_ar 0 0 1000 1034
% figure_dent_reactor_xe 0 0 1000 1046

\chapter{A Precision Neutrino Laboratory at the Spallation Neutron Source}
\chapterauthor{Jason Newby}

\chapteraffil{Oak Ridge National Laboratory, Oak Ridge, TN 37831, USA}

\chapterdoi{10.5281/zenodo.3464604}

The first observation of coherent elastic neutrino nucleus scattering (\cevns) by the COHERENT collaboration at the Spallation Neutron Source demonstrated the capability of the facility in experimental neutrino physics. 
COHERENT is now measuring the target scaling of the interaction with multiple small-scale, first-light detectors. 
These measurements are already setting new limits on nonstandard lepton-quark interactions. 
The potential of the neutrino source is fully realized with a suite of high precision neutrino measurements with ton-scale instruments. 
The neutrino flux is known within 10\% and will present a floor in the uncertainty for detectors now being considered for a precision program. 
The collaboration plans to deploy a heavy water detector to directly measure the neutrino flux at the SNS using the well-known \nue-deuteron charged current cross-section. This interaction, known to a few percent, is detected via Cherenkov light from the produced electron.  
A sufficient number of interactions will be recorded to achieve this statistical precision within two year in a ton-scale detector with an optimized design to suppress backgrounds. 
This precision measurement will ensure that the greatest impact is achieved from the planned suite of more massive \cevns targets including argon, germanium, and sodium.
\chapter{The COHERENT NaI{[Tl]} Detector}
\chapterauthor{Sam Hedges}

\chapteraffil{Duke University, Durham, NC 27708, USA}
\chapteraffil{Triangle Universities Nuclear Laboratory, Durham, NC 27708, USA}

\chaptercollab{COHERENT}

\chapterdoi{10.5281/zenodo.3462345}

COHERENT is deploying a NaI{[Tl]} detector to the SNS for measuring \cevns recoils off \iso{23}{Na} nuclei. 
The measurement of the \cevns cross section on a light nuclei will help verify its $N^2$ scaling with neutron number and provide further tests of the standard model. 
A 185-kg prototype detector has been deployed to the SNS and acquiring data since 2016. 
This prototype is providing an \emph{in situ} measurement of low-energy backgrounds as well as for studying the electron neutrino charged-current interaction on \iso{127}{I}. 
The current design parameters (subject to change) of the detector are listed below:
\begin{itemize}
\item Mass: between 2079 kg and 3388 kg. Detector designed to be modular, with final mass determined by digitization costs and space constraints. Additional mass available for future deployment.
\item Shielding: 7'' of water on all sides surrounding detector, followed by 2'' of lead outside the water.
\item Distance from target to center of array: $\sim$21 m.
\item Threshold: 3 keVee.
\item Steady-state backgrounds: 200-500 counts/keV/kg/day in the energy ROI, before background reduction from beam timing. 
\item Energy resolution:  $\textrm{FWHM}/E = \sqrt{\alpha E^2 + \beta E + \gamma}$, with $\alpha=0.00153$, $\beta=1.86$, $\gamma=1.93 \times 10^{-7}$, and $E$ in keVee.
\item Quenching factor assumption: flat, 19.65\% $\pm$ 3.76\% in the ROI.
\end{itemize}
\chapter{The CONUS Coherent Reactor Neutrino Scattering Experiment}
\chapterauthor{Manfred Lindner}

\chapteraffil{Max-Planck-Institut f\"ur Kernphysik, Postfach 103980, D-69029 Heidelberg, Germany}

\chapterdoi{10.5281/zenodo.3464513}

Coherent elastic neutrino nucleus scattering (\cevns) has been predicted since 1973 and was first observed in 2017 with neutrinos from pion decay at rest. 
CONUS aims at detecting \cevns with low energy reactor anti-neutrinos. 
It uses novel Germanium detector technology and a virtual depth shield for operation at shallow depth only 17 meters away from the core of a multi GW power reactor. 
The talk will describe the experiment, the latest results and the potential of future detectors of this kind. 
\chapter{{\color{gray}CONNIE}}
\chapterauthor{Juan Estrada}

\chapteraffil{Fermi National Accelerator Laboratory, Batavia, IL 60510, USA}

\chapterdoi{10.5281/zenodo.3464481}
\chapter{MINER -- A Reactor Coherent Neutrino Scattering Experiment to Search for Sterile Neutrinos and Non-Standard Interactions}
\chapterauthor{Rupak Mahapatra}

\chapteraffil{Mitchell Institute for Fundamental Physics and Astronomy, Department of Physics and Astronomy, Texas A\&M University, College Station, TX 77845, USA}

\chapterdoi{10.5281/zenodo.3464520}

The Mitchell Institute Neutrino Experiment at Reactor (MINER) is a reactor based experiment at Texas A\&M university that combines well-demonstrated low-threshold cryogenic detectors developed for the SuperCDMS dark matter search with a unique megawatt research reactor that has a movable core providing meter-scale proximity to the core. 
The low-threshold detectors ($\approx 100$ eV recoil energy) will allow detection of coherent scattering of low energy neutrinos that is yet to be detected in any reactor experiment. 
These high resolution detectors, combined with a movable core, provide the ideal setup to search for short-baseline sterile neutrino oscillation by removing the most common systematic in current experiments, the reactor flux uncertainty. 
Very short baseline oscillation will be explored as a ratio of rates at various distances, with expected SM rates and known scaling of background. 
Hence MINER will be largely insensitive to absolute reactor flux. 
Additionally, low variation in a MW research reactor power combined with meter-scale proximity to the core provides much better systematics compared to a GW power reactor, where the typical detector to core distance is of the order of 30 meters or higher resulting in similar neutrino flux incident on a detector. 
Utilizing multiple targets (Ge/Si) allows for detailed understanding of the signal and backgrounds in the experiment. 
Precise understanding of the background is important for searches of Non Standard Interactions (NSI) through a small additional signal.

Phase-1 of the MINER experiment is already operational as a demonstration experiment with a 2-kg (4-kg maximum capacity) payload at a distance of approximately 4.5 m from the reactor core, that would provide a signal rate approaching 1000 events per year and a target background of 100-1000 counts/keV/kg/day (DRU). 
Phase-2 of the MINER experiment experiment will have a 20 kg payload (inside a 30-kg infrastructure), using a recently purchased cryogen-free refrigerator. 
The operational 2-kg demonstration phase provides an excellent opportunity to design the full MINER experiment with 10x larger payload, 10x higher flux due to proximity to core and 10x lower background due to hermetic passive and active shielding. 
The sensitivity to \cevns will improve by at least two orders of magnitude, allowing for precisions tests of eV-scale sterile-$\nu$, Non-Standard Interactions and neutrino magnetic moment.

The MINER experiment aims to become the first experiment to measure \cevns at a reactor and may open windows to much exciting new physics of immediate interest:

\begin{description}
	\item[Precision \cevns] with high statistical ($\sim$1,000 events/kg/year and systematic precision using low-threshold semiconductor detectors (Ge/Si) at close proximity ($\approx$2 m) to core and passive/active shielding. Our measurements would be an independent confirmation of recent observations by the COHERENT Collaboration and provide important complementarity to the prompt muon neutrino signal at SNS to constrain NSI.
	\item[Search for sterile neutrinos] as a possible deficit in predicted Standard Model rates using a precisely movable core. MINER's very short baseline (1--10 m) search provides important complementarity to the PROSPECT non-coherent (IBD process) search.
	\item[Search for light and heavy $Z^\prime$ and NSI.] For light $Z^\prime$ down to a mass scale of 1 MeV, the sensitivity can improve upon that of fixed target and atomic parity violation experiments. For heavy $Z^\prime$ up to a mass scale of 4 TeV, the sensitivity is competitive with and complementary to LHC searches. Due to different flavor composition, the sensitivity to light and heavy $Z^\prime$ and NSI will be complementary to that of the COHERENT experiment.
\end{description}
\chapter{Precision measurement of \cevns (Ge PPCs @ COHERENT) }
\chapterauthor{Juan I. Collar}

\chapteraffil{Enrico Fermi Institute and Kavli Institute for Cosmological Physics, University of Chicago, Chicago, IL 60637, USA}
\chapteraffil{Department of Physics, University of Chicago, Chicago, IL 60637, USA}

\chapterdoi{10.5281/zenodo.3464440}

The state-of-the-art in p-type point contact (PPC) germanium detectors has reached a level of maturity sufficient to envision their use for \cevns measurements at both spallation and reactor sources. 
Crystals in the 3--4 kg mass range are presently under development with noise characteristics sufficient to provide a 150 eV ionization ($\sim$600 eV recoil) energy threshold, with high ($>$80\%) signal acceptance at threshold. 
The intrinsic radiopurity of these devices allows to reach levels of background in the few counts/keV kg day, for sites having only a modest (few m.w.e.) overburden.

\chapter{New Constraints on the matter potential from global analysis of oscillation data}
\chapterauthor{Ivan Martinez-Soler}

\chapteraffil{Fermi National Accelerator Laboratory, Batavia, IL 60510, USA}
\chapteraffil{Northwestern University, Evanston, IL 60208, USA}

\chapterdoi{10.5281/zenodo.3464581}

The neutrino evolution is a long-standing problem in particle physics
since many decades ago. In the light of the latest global oscillation
analysis, we are entering into the precision
era~\cite{Esteban:2018azc}\footnote{Editor's note: this reference has been updated to reflect the published version of the work originally cited only as a preprint.}. The description of the neutrino evolution
in matter is crucial for the determination of most of the remaining
uncertainties in the neutrino evolution:
\begin{itemize}
\item The measurement of the mass ordering driven by
  Super-Kamiokande~\cite{hayato:2018} depends on the measurement of
  the 1-3 mixing resonance in atmospheric neutrinos crossing the
  Earth's mantle with $E_{\nu}\sim 6$~GeV;
\item the phase that violates the CP symmetry in the lepton sector is
  measured by Super-Kamiokande~\cite{hayato:2018} in the interference
  region between the 1-2 ($E_{\nu}\sim 0.1$~GeV) and 1-3 mixing
  resonances;
\item the two issues that contribute to the tension in the
  determination of the solar mass parameter~\cite{Esteban:2018azc} are
  the matter effects introduced by the Earth over the solar neutrinos,
  and the turn up of the solar spectrum in the low energy region where
  the solar matter effect dominates.
\end{itemize}
In the presence of non-standard interactions (NSI) of the neutrino
with the matter, their evolution and therefore the determination of
the oscillation parameter will be altered. In this talk, we are going
to discuss our knowledge of the size and flavor structure of NSI by a
global fit of oscillation data~\cite{Esteban:2018ppq}, considering a
general neutral current neutrino interaction with quarks. We assume
that the lepton-flavor structure of the new interactions is
independent of the quark type. The results have been obtained using
all the available data from oscillation experiments alone and in
combination with the results on coherent neutrino-nucleus scattering
from the COHERENT~\cite{coherentScience2017} experiment. In our analysis,
we study the robustness of the three neutrino mixing scenario in the
presence of NSI, and the LMA-D solution. As a result, we also derive
new bounds of the non-standard couplings to up and down quarks. The
results obtained are robust under the broad spectrum of up-to-down
strengths found in the neutrino propagation along the Sun and the
Earth.
\chapter{Light sterile neutrinos: the 2018 status}
\chapterauthor{Stefano Gariazzo}

\chapteraffil{Instituto de F\'isica Corpuscular (CSIC-Universitat de Val\`encia), Paterna (Valencia), Spain}

\chaptercoauthor{C. Giunti, M. Laveder, Y.F. Li}

\chapterdoi{10.5281/zenodo.3462638}

In the recent years, sterile neutrinos with a mass around 1~eV have been studied as a possible solution for the Short-BaseLine (SBL) neutrino oscillation anomalies, which include results from LSND \cite{Aguilar:2001ty} and MiniBooNE \cite{Aguilar-Arevalo:2018gpe}, from GALLEX and SAGE \cite{Giunti:2010zu} and from a number of reactor antineutrino experiments \cite{Mention:2011rk}.
These experimental measurements cannot be explained in the context of the standard three neutrino oscillations.
The current status of the search of such light sterile neutrino has been reviewed using all the available appearance and disappearance data in SBL experiments.
Muon (anti)neutrino disappearance as constrained mainly by the IceCube and MINOS/MINOS+ experiments is substantially in tension with the observation of electron (anti)neutrinos appearance in a flux of muon (anti)neutrinos, as observed by LSND and MiniBooNE, when also the electron (anti)neutrino disappearance results are taken into account \cite{Gariazzo:2017fdh}.
From this latter channel, however, we have the first model-independent indications \cite{Gariazzo:2018mwd} in favor of active-sterile neutrino oscillations, thanks to the observations from the NEOS \cite{Ko:2016owz} and DANSS \cite{Alekseev:2018efk} experiments.
The two collaborations aim at measuring the reactor antineutrino flux at different distances (between 10 and 25~m) in order to distinguish the effect related to a global normalization, which does not depend on the distance at which the measurement is performed, from the one due to neutrino oscillations, which instead varies with the baseline.
In the incoming years, these and other currently running experiments will use standard techniques to test the current best-fit parameters and probe the signal observed by LSND and MiniBooNE, to definitely confirm or rule out the existence of a light sterile neutrino.
Meanwhile, the first \cevns experimental results have also been employed to derive bounds on the active-sterile neutrino.
While at the moment these probes are not competitive with the above mentioned constrains \cite{Kosmas:2017tsq}, \cevns will play a role in this game in the future \cite{Canas:2017umu}: studies show that experiments based on coherent scattering will be extremely useful to test neutrino oscillations at reactors, at very small distances (possibly down to 1-3~m): these experiments, therefore, will be perfect probes for active-sterile neutrino oscillations.

\chapter{Complementarity Short-Baseline Neutrino Oscillation Searches with \cevns}
\chapterauthor{Joel Walker}

\chapteraffil{Department of Physics, Sam Houston State University, Huntsville, TX 77341, USA}

\chaptercoauthor{B. Dutta, J. Dent}

\chapterdoi{10.5281/zenodo.3464701}

Anomalies in the expected magnitude and spectrum of neutrino flux have been pointed out for several years in reactor ($\bar{\nu}_e$ deficit) and Gallium ($\nu_e$ deficit) data.
Newer reactor (Daya Bay, DANSS, NEOS) analyses take ratios of observations at different baselines in order to remove dependence upon the flux normalization and intrinsic spectral shape; inclusion of a sterile then improves the goodness of fit at around $3\sigma$ preference.
Recently, the accelerator-based MiniBoone experiment has presented results (${\nu}_e$ appearance within a $\nu_\mu$ beam) consistent with anomalies observed previously by LSND.  Detection is flavor-sensitive, with $E_\nu \simeq$~500~MeV and $L\simeq$~0.5~km.
Neutrino 4 employs segmented IBD detection at a MW research reactor with $L=6\text{--}12$~m.  IBD is flavor sensitive and fully reconstructs the neutrino energy, allowing for ``coherency'' of an oscillation signal over may cycles in $L/E$.  A relative preference ($\Delta \chi^2$) for oscillation is reported at the level of $3\sigma$.
The various anomalies are generally consistent within ``types'' (when multiple experiments exist), although it is difficult to reconcile them across types with simple models.

Various CEvNS experiments are well-positioned to probe possible connections of a short-baseline neutrino oscillation effect to existing anomalies.
New physics will be most visible to CEvNS when it impacts the expected event distribution shape (e.g. for light mediators, magnetic moment, and steriles), rather than only the rate (e.g. for heavy \zprime).
Large statistics associated with the CEvNS coherency enhancement can allow for precision discrimination.
Considerable complementarity in the flavor and mass space is possible by a combination of experimental efforts.
The CEvNS neutral current touches all flavors.  Prompt/delayed signal discrimination for COHERENT at the SNS, together with reactor data, and application of the unitarity constraint provides enough information to independently constrain this multi-flavor system at the matrix element level.
The SNS uses stopped pions to produce an isotropic prompt monochromatic $\nu_\mu$ with $E \simeq 30$~MeV, and secondary isotropic delayed $\bar{\nu}_\mu$ and $\nu_e$ with calculable energy spectra.  This high energy enhances the cross section (as a square) and allows for comparatively simple detectors (threshold requirements are also quadratic) that scale well to high mass.  Background control is good, enhanced by timing information.  However, the neutrino flux is lower than typical reactor facilities by around 5 magnitude orders.
At reactors, neutrino flux is extraordinarily high ($10^{12-13}~{\rm cm}^{-2}{\rm s}^{-1}$).  But, backgrounds are challenging and detectors (e.g. high-voltage Ge/Si with transition-edge phonon sensors) must be carefully designed for sensitivity to soft recoil.  Additionally, the reactor spectrum is widely spread across energies in the few MeV range (although the shape is reasonably well known), which leads to dispersion of an oscillation signal beyond 1-2 cycles.  Binning in the recoil, and the extraction of independent bins at ultra-low recoil, can help greatly.  An advantage is that this spread in energies is effectively equivalent to scanning multiple length baselines simultaneously.  Simplifications include lack of flavor ambiguity and ability to neglect the nuclear form factor (coherency is fully maintained).  In the future, directional detection could resolve the event-by-event neutrino energy ambiguity. 
Additional important experimental complementarities include diversity in the range of $L/E$ deployments and nuclear target materials.
\chapter{Reactor fluxes for \cevns}
\chapterauthor{Patrick Huber}

\chapteraffil{Center for Neutrino Physics, Virginia Tech, Blacksburg, VA 24061, USA}

\chapterdoi{10.5281/zenodo.3462685}

Reactor neutrinos have played a central role in neutrino physics since
their first detection by Reines and Cowan~\cite{Cowan:1992xc}; since
then every experiment needed to have some understanding of emitted
antineutrino flux and energy distribution stemming from a
reactor. Much work has been done for the antineutrino yield above the
inverse beta decay (IBD) threshold~\cite{mueller2011a,huber2011a},
but little work for lower energies. \cevns being a threshold-less
reaction therefore puts up new challenges, but of course for now we
await experimental progress towards the detection of reactor
antineutrinos using \cevns and it will be a while before antineutrinos
below IBD threshold will be detected, since these correspond to recoil
energies of $\sim100$\,eV or less. Standard lore is that neutron
captures play a negligible or at best percent-level role for
antineutrino yields~\cite{huberJaffke2016} above IBD threshold. This is
different at low energy, with the most abundant reaction being
$^{238}\mathrm{U}+n\rightarrow~^{239}\mathrm{U}$, which then leads to
two beta decays with neutrino energies of up to 1.2\,MeV. The
antineutrino yield from this reaction exceeds the one of all fission
fragments by an order of magnitude at 1\,MeV. Other reactions have
been pointed out where neutrons capture on structural materials in the
reactor core~\cite{alum}. A detailed survey of nuclear data bases
certainly will reveal more relevant isotopes and then the question of
how well known the neutron capture cross sections actually are will
arise. Another interesting wrinkle,
are the potential contributions from $\beta^+$ decays which would yield
neutrinos instead of antineutrinos. Neutron captures do not represent
a fundamental issue, but care needs to be taken to understand the
specific reaction rates in a specific core with its specific core
inventory; at low energies reactors start to become individual
neutrino sources whose details matter. Also, at low energy the
half-lives of antineutrino emitters increase significantly and thus
many isotopes will be not in equilibrium, as a consequence the
operational history of a reactor becomes important as well and
instantaneous thermal power no longer is a direct predictor for the
antineutrino flux. Increased collaboration with and input from nuclear
engineers and reactor operators will be needed to address these
issues.

 This work was supported by the U.S. Department of Energy
 under award number \protect{DE-SC0018327}.

\chapter{{\color{gray}Exploring New Roles for \cevns and Neutrinos}}
\chapterauthor{Bernadette Cogswell}

\chapteraffil{The University of Manchester, Manchester, M13 9PL, UK}

\chapterdoi{10.5281/zenodo.3464436}

\chapter{NU-CLEUS: Exploring \cevns at low energies with cryogenic calorimeters}
\chapterauthor{Raimund Strauss}

\chapteraffil{Technical University of Munich, D-85748 Garching, Germany}

\chaptercollab{NU-CLEUS}

\chapterdoi{10.5281/zenodo.3464650}

The NU-CLEUS experiment is a new experiment \cite{Strauss:2017cuu} to explore coherent-neutrino nucleus scattering (\cevns) \cite{freedman74, drukierStodolsky} at a nuclear power reactor.
Recent results from a prototype gram-scale cryogenic calorimeters (gramCC) \cite{Strauss:2017cam} operated at the Max-Planck-Institute for Physics (MPP), opened a new window to neutrino physics at unprecedentedly low energies. 
An energy threshold for nuclear recoils of $\left(19.7\pm0.9\right)$~eV was reached, which is one order of magnitude lower than previous results of macroscopic cryogenic detectors \cite{angloher2016cresst} and a factor of 50~--~100 lower than the state-of-the-art germanium detector technology based on ionization technology. 
This breakthrough enables a rich physics program to study the fundamental properties and interactions of neutrinos, to perform precision tests of the electroweak theory as well as nuclear and reactor physics. NU-CLEUS aims for the exploration of \cevns at the low-energy frontier which opens the door for new physics beyond the Standard Model of Particle Physics. \\
%
% bounding boxes for images
% figure_strauss-Detector 0 0 927 492 (or, from designer looking at JPG: 0 0 1544 820
% figure_strauss-Background 0 0 757 538
%
\begin{figure}[htbp]
\centering
	\begin{subfigure}[h]{0.49\textwidth}
		\includegraphics[width=\textwidth, bb = 0 0 1544 820]{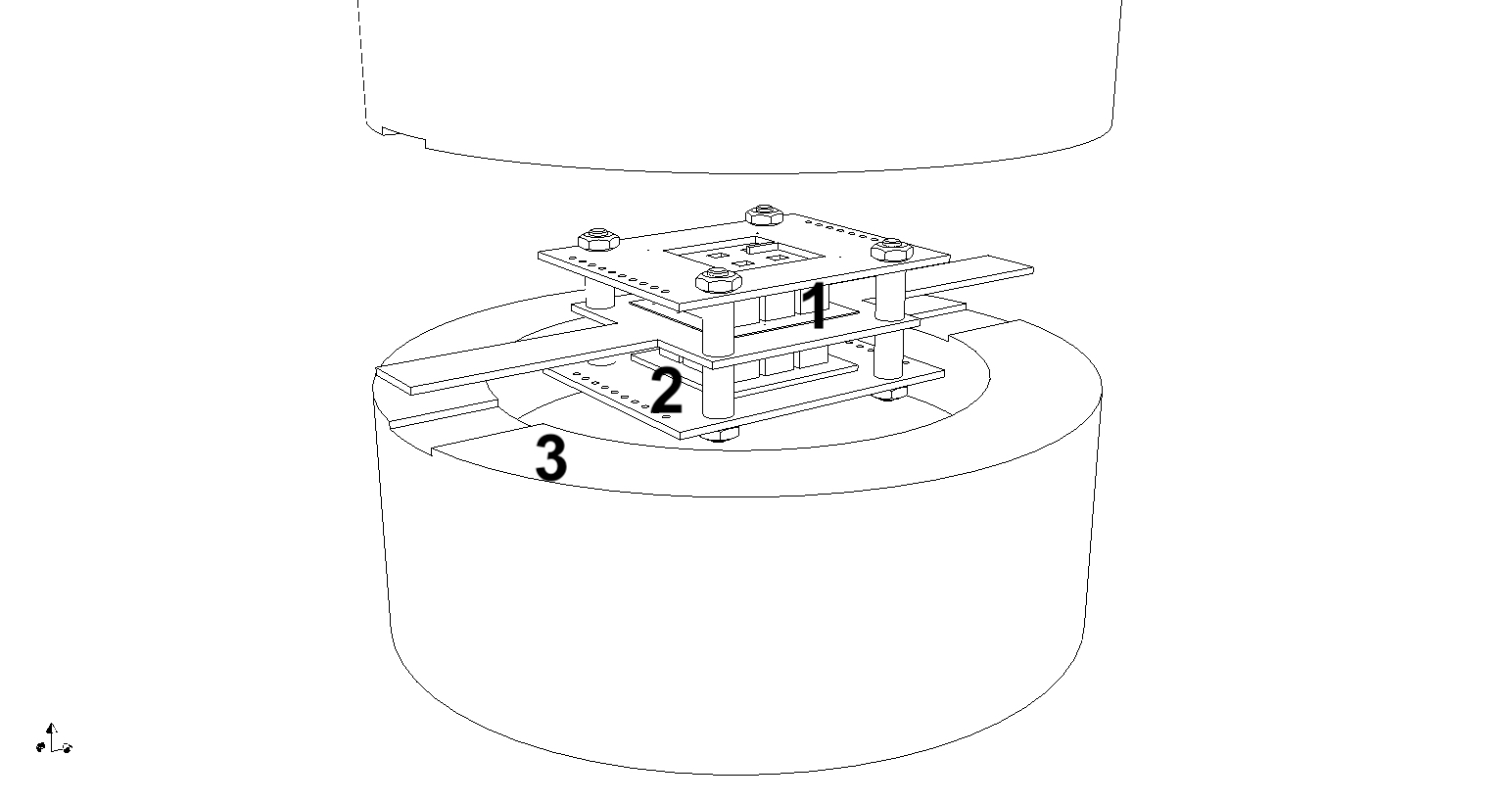}
	\end{subfigure}
	\begin{subfigure}[h]{0.49\textwidth}
		\includegraphics[width=\textwidth, bb = 0 0 757 538]{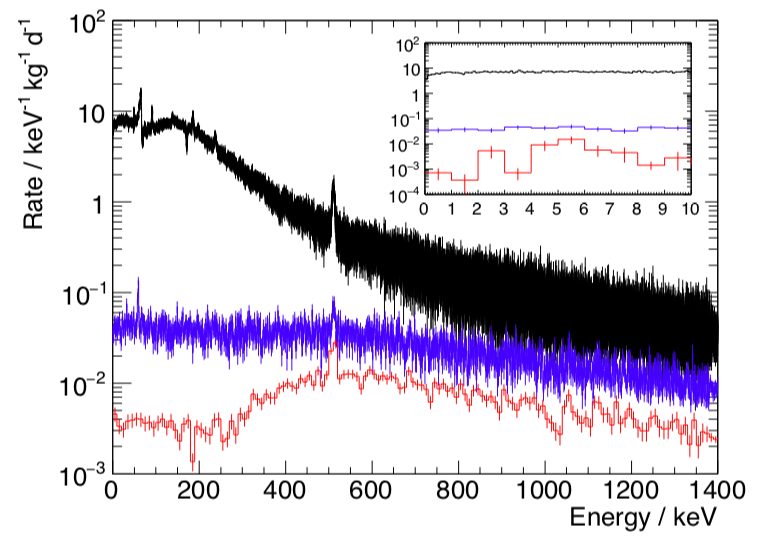}
	\end{subfigure}
	\caption{Left: Technical drawing of the NU-CLEUS prototype detector. 1) target, 2) inner veto, 3) outer veto. Right: MC simulation of the expected energy deposit in case of a background similar to the remaining one in the Dortmund Low Background facility. Black: Without any veto, Blue: in case of a passive outer veto, red: in case of an active outer veto with a threshold of 1 keV. Figure adopted from \cite{Strauss:2017cuu} and references therein.}
\label{fig:strauss1}
\end{figure}
The gramCCs will be operated within a fiducial-volume cryogenic detector, a promising new concept developed for NU-CLEUS, that is suited for an above ground operation at significantly suppressed background levels. 
It consists of three subsets of cryogenic calorimeters -- the outer and the inner veto and a neutrino target -- all operated at mK temperatures (see Fig. \ref{fig:strauss1} left). 
Located at a nuclear power reactor this detector has the potential to achieve a signal to background ratio of up to $10^3$, a unique situation in neutrino physics \cite{Strauss:2017cuu} (see Fig. \ref{fig:strauss1} right). 
This will enable a rapid observation of \cevns within 1~--~2 weeks with a total detector mass of 10 g. 
After a measuring time of only 10 weeks we will exceed the precision of today's results of COHERENT \cite{coherentScience2017} and overcome its systematic uncertainties. 
\chapter{The Very Near Site at Chooz - a New Exerimental Hall to Study \cevns}
\chapterauthor{Victoria Wagner}

\chapteraffil{IRFU, CEA, Universit\'{e} Paris Saclay, F-91191 Gif-sur-Yvette, France}

\chaptercoauthor{T. Lasserre, A. Langenk\"amper, C. Nones, J. Rothe, R. Strauss, M. Vivier, A. Zolotarova}

\chapterdoi{10.5281/zenodo.3464679}

The Very-Near-Site (VNS) is a very promising new experimental site for future experiments to study coherent elastic neutrino nucleus scattering (\cevns). 
With a baseline of 72\,m and 102\,m, respectively, the VNS is located in the close proximity of the two reactor cores of the Chooz nuclear power plant in France, each running at a nominal thermal power of 4.25\,GW$_{\mathrm{th}}$. 
The expected anti-neutrino flux at the VNS is of the order of $10^{12}\,\mathrm{cm}^{-2}\cdot\mathrm{s}^{-1}$.  
Any experimental setup at the VNS is restricted both in weight and volume. 
First muon attenuation measurements indicate a shallow overburden of 3\,m.w.e.:  \cevns experiments located at the VNS face a high cosmic muon rate of the order of 100\,$\mathrm{m}^{-2}\cdot\mathrm{s}^{-1}$ and a potentially challenging muon-induced neutron background. 

The NU-CLEUS detector concept \cite{Strauss:2017cuu} provides a suitable technology for a next generation \cevns experiment at the VNS. 
Thanks to the unprecedented low energy threshold of $\leq$\,20\,eV$_{\mathrm{nr}}$ \cite{Strauss:2017cam}, a strong \cevns signal is expected even for gram-scale target masses. 
To reduce muon-induced backgrounds, the experimental volume containing the NU-CLEUS detectors will be surrounded by a passive shielding which will be complemented with an active muon-veto. 
With the fast rise-time of the NU-CLEUS detectors, the muon-induced dead-time will stay below a few percent, even for large surfaces of the active muon-veto operated at shallow overburden. 
Additional and complementary ways to fight the backgrounds are being investigated. As such, the BASKET \cite{basket:online} R\&D program seeks to develop detectors which could achieve an in-situ neutron background characterization. 
In a first phase, a 10-g version of the NU-CLEUS detector is planned to be installed and commissioned at the VNS in 2020.
\chapter{The Ricochet Experiment} \label{chap:ricochet}
\chapterauthor{J.A. Formaggio}

\chapteraffil{Massachusetts Institute of Technology, Cambridge, MA 02139, USA}

\chaptercollab{Ricochet}

\chapterdoi{10.5281/zenodo.3464500}

Ricochet is a bolometer-based CENNS experiment aimed at measuring neutrinos created from the fission process deep within the reactor core.  In the first phase of its deployment, Ricochet aims to deploy a 1 kilogram load to measure the nuclear neutrino flux.  Ricochet leverages two cryogenic technologies as part of its measurement program:

\begin{enumerate}

\item \protect{{\it Germanium Semi-conductors}}:   In semiconductor bolometers, the rejection between backgrounds and CENNS-signal events will be achieved thanks to the double measurement of the heat and ionization energies, which ratio depends on the nature of the interacting particle: gamma- or beta-induced electronic recoils (electromagnetic interactions), CENNS- or neutron-induced nuclear recoils (lattice interactions). The goal is to reach \protect{$\sim$}10~eV (RMS) energy resolution in heat and \protect{$\sim$}20~eV (RMS) resolution in ionization to provide a rejection power of \protect{$10^3$} down to the energy threshold. To reach such outstanding background rejection to all sorts of electromagnetic backgrounds, two key features have to be met: i) low-capacitance (\protect{$\sim$}10 pF) Fully Inter-Digited (FID) electrodes, as first introduced by the EDELWEISS collaboration~\protect{\cite{Armengaud_2017}}, thanks to which events happening near the surface (within ~100 \protect{$\mu$}m) can be tagged as such and be rejected while providing excellent charge collection for bulk events; ii) $\sim$20~eV eVee ionization energy resolution (RMS) per electrode, which is five times better than the best resolution achieved so far in such massive cryogenic bolometers~\protect{\cite{PHIPPS2019181}}. 

\item \protect{{\it Metallic Zinc-Superconductors}}: In Zn-detectors, due to the vanishing quasiparticle-phonon coupling in superconducting metals below \protect{$\sim$}100 mK and the difference in thermalization processes between electronic recoil backgrounds and CENNS-induced nuclear recoils, we expect vastly different heat pulse shapes between these two populations of events. From preliminary simulations, assuming a quasiparticle recombination rate 5 times longer than the phonon thermalization rate, we expect a background rejection power of \protect{$10^3$} down to the energy threshold. Indications that such behavior may indeed exist in such target medium has been documented by the MARE collaboration in their measurements using superconducting rhenium and alpha particles (\protect{$\sim 5$} MeV)~\protect{\cite{Cosulich1993, Booth_1996}}.
\end{enumerate}

In January 2019, the strong synergy between the R\&D programs of Ricochet and EDELWEISS resulted in the successful demonstration a 55 eV energy threshold on a 33.4-g Ge bolometer operated from a surface lab (IPNL).  Thanks to the outstanding performance and stability of the detector, the best surface-based Dark Matter limit on both Strongly and Weakly Interacting Massive Particles down to 600~MeV/c$^2$ has been derived~\cite{Armengaud:2019kfj}. Nowadays, the main R\&D focus is  dedicated to demonstrating the rejection capabilities of the electromagnetic backgrounds down to the energy threshold. To that end, a first version of a HEMT-based preamplifer developed by the IPNL group is being tested, and new electrode designs are being developed in parallel. An intermediate goal of a 30-g scale Ge bolometer combining a 20 eV heat energy resolution (RMS) together with a 50 eV ionization energy resolution (RMS) is planned for the end of 2019.

At the end of 2019, the two technologies will be scaled to multiple targets for a total target mass of 1 kg.  The two detector types, {\it CryoCube} and {\it Q-Array} are briefly described below.

\begin{itemize}
    \item The CryoCube: it consists of an array of $3\times 3 \times 3 = 27$ single 30-g Ge and Zn highly performing cryogenic detectors. The crystals will be packed together following a Rubik's cube like topology in a $8 \times 8 \times 8$~cm$^3$ radio-pure infrared-tight copper box suspended below the mixing chamber with its dedicated cryogenic suspension system~\cite{Maisonobe:2018tbq}, and its cold front end electronics in close proximity of the detectors. Each single crystal is designed to reach a $\mathcal{O}$(10) eV energy threshold and a $10^3$ electromagnetic background rejection power down to the energy threshold. This detector array is fully funded by the CENNS - ERC starting grant.
    
    \item The Q-Array:  it consists of an array of 8 or 16 superconducting 40-gram zinc cubes.  Each unit will be read out by a transition-edge sensor and the signal feed into a microwave resonant SQUID array (uMUX), allowing the signals from multiple detectors to be read out by a single feed line.  The uMUX array operates at frequencies near 7 GHz, with each channel specifically tuned to a corresponding resonant frequency set by the capacitance of the transmission line. A prototype SQUID array has been designed and produced by Lincoln Laboratories at MIT and is currently undergoing testing.  Initial results show excellent quality factors ($Q \simeq 50,000$), and results on gain and noise performance will be available this spring.  Further amplification is applied using a tunneling-wave parametric amplifier (TWPA), also developed at Lincoln Laboratory.  It is expected to bypass the need for a HEMT amplifier.  The Q-Array geometry is expected to easily integrate with the Cryocube support structure, allowing for the two technologies to co-inhabit the same cryogenic space and share shielding options.    
\end{itemize}

The collaboration is currently determining the optimal location for deployment of the CryoCube and Q-Array demonstrators.  Table~\ref{tab:rates} shows the expected rate from a 1 kilogram target with our target energy threshold of 50 eV at several sites considered by the collaboration.

\begin{table}[htp]
    \centering
    \begin{tabular}{|l|c|c|c|}
    \hline
    Reactor (Location) & Thermal Power & Distance & Event Rate (per day) \\
    \hline 
MITR (USA) & 5.5 MW  & 4 meters & 7.4 \\
ILL (France)\footnote{Likely site for our detector deployment.} & 58.3 MW  & 8 meters & 19.5 \\
Double Chooz (France) & 4250 MW  & 80 meters & 14.3 \\
    \hline
    \end{tabular}
    \caption{The event rates expected at various reactor sites assuming a 1 kg germanium or zinc bolometer with 50 eV recoil energy threshold.}
    \label{tab:rates}
\end{table}

A decision as to which location will be a first deployment will be completed in 2019.

\chapter{The Cryocube Detector Array for Ricochet}
\chapterauthor{Dimitri Misiak}

\chapteraffil{Univ Lyon, Universit\'e Lyon 1, CNRS/IN2P3, IPNL-Lyon, F-69622 Villeurbanne, France}

\chaptercollab{Ricochet}

\chapterdoi{10.5281/zenodo.3464594}

The Cryocube project, being part of the Ricochet experiment, aims for a percentage-level precision measurement of the CENNS process to probe various exotic physics scenarios. It consists in a cubic compact array of cryogenic detectors with the following specifications:

\begin{itemize}
	\item a very low energy threshold of $\mathcal{O}(10)$eV on the phonon heat signal,
    \item an electromagnetic background rejection of at least $10^3$,
    \item a total target mass of 1kg divided between 27 crystals of $32$g,  
    \item two complementary target elements: germanium and zinc.
\end{itemize}
Investigation on the thermal sensor technology (NTD Germanium, NbSi TES) and detector thermal modelization is ongoing with a first prototype that achieved a $55$eV of energy threshold, within the EDELWEISS R\&D program. The event discrimination is realized in semiconductor germanium crystals with HEMT-based ionization readout to reach $\mathcal{O}(10)$eV in ionization resolution, and in superconducting zinc crystals with heat pulse shape discrimination. An accurate low-energy measurement of the Quenching factor will be conducted using an in-situ neutron calibration based on the multiple detector coincidence. The installation of the Cryocube in a dry cryostat with shielding and already proven vibration-decoupling strategy is planned within three years near an optimal nuclear reactor. After an exposure of $1~\text{kg}\cdot\text{year}$,  a percentage-level precision measurement of the CENNS process will be delivered by 2024.
\chapter{BULLKID - Bulky and low-threshold kinetic inductance detectors}
\chapterauthor{Marco Vignati}

\chapteraffil{INFN, Sezione di Roma, P.le Aldo Moro 2, 00185 Roma, Italy}

\chaptercollab{BULLKID proto-\!}

\chapterdoi{10.5281/zenodo.3464677}

BULLKID is an R\&D for new experiments on sub-GeV dark matter and coherent neutrino-nucleus scattering which leverages the sensitivity and high multiplexing capability of superconducting Microwave Kinetic Inductance Detectors (MKIDs~\cite{daynature}) to reach low energy thresholds ($<100$\,eVnr) and high target masses ($\sim1$\,kg). The project, funded by INFN in Italy, started in January 2019 and includes collaborators from INFN-Rome, INFN-Ferrara, INFN-Genova, IFN-CNR, CEA, Institut N\'eel and Zaragoza U. 

\begin{figure}[htbp]
\begin{center}
\includegraphics[width=0.9\textwidth, bb = 0 0 790 358]{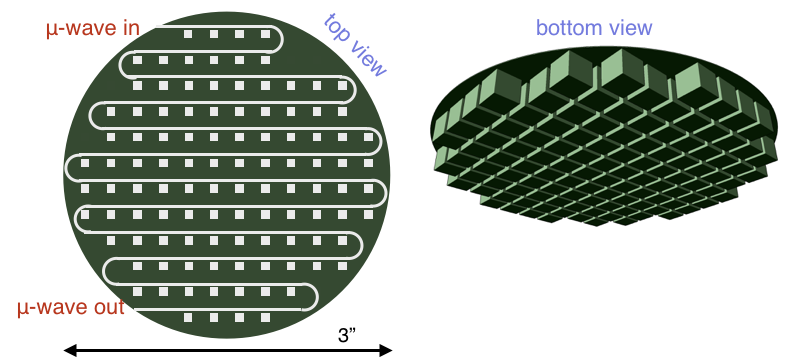}
\caption{The proposed BULLKID layout. Left: around 100 MKID sensors are deposited on a 5 mm thick, 3'' diameter silicon wafer and coupled to a single feedline for multiplexing. Right: the wafer is diced from the bottom to obtain independent 5x5x5 mm$^3$ voxels acting as particle absorbers.}
\label{fig:bullkid}
\end{center}
\end{figure}

The  detector is based on the technology developed within CALDER~\cite{calder}: particles interact in a silicon substrate and generate athermal phonons that scatter trough the substrate until they are absorbed and generate a signal in aluminium~\cite{calderal} or aluminium-titanium~\cite{calderaltial} MKIDs deposited on the surface. While CALDER is using 2x2\,cm$^2$ or 5x5\,cm$^2$, 300\,$\mu$m thick, silicon substrates as photon absorbers, the Dark Matter or neutrino target of BULLKID will consist in a 5x5x5 mm$^3$ silicon voxel. To exploit the MKID multiplexing several voxels will be carved from a single 5 mm thick silicon ÒwaferÓ with a diameter of 3'': one side of the wafer will host the lithography, with a single feedline running through all the MKIDs (Fig~\ref{fig:bullkid}, left); the opposite side will be cut into a square grid of ~5 mm pitch with a 4.5 mm dice depth, so as to obtain almost cubic voxels, and leave the surface hosting the MKIDs intact (Fig~\ref{fig:bullkid}, right). In this way the phonons produced in a voxel will be isolated, and absorbed by a single MKID to improve the signal to noise ratio. This geometry also ensures an efficient background reduction, as multiple-voxel events can only be generated by cosmic rays or natural radioactivity and not by neutrinos or dark matter particles. Around 100 voxels of 0.3\,g each can be obtained from a 3'' wafer, for a total active detector mass of 30\,g. In a future experiment with higher target mass several wafers could be stacked and read independently to reach higher target masses.
%
% bounding box for image
% 0 0 790 358
%
\chapter{Towards 10-kg Skipper CCD detectors}
\chapterauthor{Javier Tiffenberg}

\chapteraffil{Fermi National Accelerator Laboratory, Batavia, IL 60510, USA}

\chapterdoi{10.5281/zenodo.3464670}

The newly developed Skipper-CCD sensor has a natural and immediate application for the detection of low energy neutrino interactions through the recently observed Coherent Elastic Neutrino-Nucleus Scattering (\cevns) process. 
The first working instrument using Skipper-CCD sensors was produced in 2016 at Fermilab in collaboration with the LBNL MicroSystems Lab. 
This system was able to unambiguously detect single ionized electrons and reach a groundbreaking 1.1 eV energy threshold, the theoretical limit of ionization detectors based on silicon (given by its band gap). 
This technological breakthrough opens a new path for miniaturized neutrino detectors by providing the capability to observe reactor neutrinos at the 1~MeV scale through the \cevns process with an unprecedented low energy threshold. 
Technical advances are required to scale up in mass and build multi-kilogram neutrino detectors.
Also, to fully profit from a mass increase, a direct measurement of the ionization efficiency of Silicon nuclei at low recoil energies is planned. 
This measurement is essential to establish the sensitivity of a silicon sensor to low energy neutrinos and has implications for other silicon based detectors such as SuperCDMS-SNOLAB and DAMIC-M. 
There is a funded R\&D path for the next 5 yrs to enable a new generation of compact detectors with unprecedented sensitivity to low energy neutrinos that will allow the exploration of their fundamental nature in the low energy regime, that is particularly interesting for new physics searches.
\chapter{The \protect{\cevns} Glow of a Supernova}
\chapterauthor{Kate Scholberg}

\chapteraffil{Duke University, Durham, NC 27708, USA}

\chaptercoauthor{A. Smith, G. Sinev}

\chapterdoi{10.5281/zenodo.3464638}

The collapse of the core of a massive star at the end of its life will produce a compact remnant such as a neutron star or black hole, in many cases a violent explosion in electromagnetic radiation and kinetic energy, and likely in all cases a brilliant burst of neutrinos over a few tens of seconds.  These supernova-burst neutrinos come in all flavors. Most are emitted quasi-thermally with energies of a few tens of MeV~\cite{Mirizzi:2015eza}.  The flavor, energy and time structure of the neutrino burst carries information about the astrophysics of the collapse, the remnant and the subsequent explosion.  It also carries information about the properties of neutrinos themselves, including information about mass hierarchy and flavor transitions within the star.

The burst of supernova neutrinos can be detected in large neutrino detectors worldwide for collapses within a few hundred kiloparsecs~\cite{Scholberg:2012id}.   Neutrino interactions with matter in the few tens of MeV range depend on flavor, energy and detector material.  Because supernova neutrino energies are less than 100 MeV, neutrinos can interact via charged-current channels only for $\nu_e$ and $\bar{\nu}_e$ flavor.  Because charged-current threshold for muon neutrinos is greater than 100 MeV, the muon and tau flavor components of the supernova burst are accessible only via neutral-current interactions.

The main existing large detector types for current and future detectors are water Cherenkov (Super-K and Hyper-K, as well as IceCube, KM3NET), liquid scintillator (LVD, Borexino, KamLAND, JUNO) and liquid argon (DUNE).  The water and scintillator detectors' primary sensitivity is the $\bar{\nu}_e$, via inverse beta decay (IBD) on free protons,
$\bar{\nu}_e+ p \rightarrow e^+ + n$.  Argon detectors are primarily sensitive to CC interactions of $\nu_e$ on $^{40}$Ar.  Some neutral-current interactions are visible via scattering on protons in liquid scintillator, the deexcitation of nuclei excited via neutral-current scattering,  and a component of elastic scattering on electrons, but these are subdominant channels.  Detection and tagging of neutral-current interactions are especially valuable in the supernova-neutrino detection game, due to the neutral current's sensitivity to the \textit{total} supernova neutrino flux. Neutral-current detection is important not only for understanding the total energy release of the supernova, but also because it enables understanding of flavor transitions within the supernova.

\cevns is a neutral-current interaction channel which may be used to measure the total neutrino flux from a supernova.  The cross section is large compared to other interactions used for supernova detection, but the produced signal is in the form of  recoiling nuclei with energies of tens of keV or less, which is well below the threshold of most supernova-sensitive neutrino detectors.  The exception is  WIMP dark matter detectors, which are now reaching tonne scale, and which will be able to observe a handful of events per tonne for a core collapse at a standard distance of 10 kpc (e.g.,~\cite{Lang:2016zhv}.) 

A new idea presented in this talk is to exploit \textit{kilotonne-scale} underground detectors for \cevns detection, to determine the total neutrino flux over the supernova burst.   The few-keV \cevns recoils are invisible in water Cherenkov detectors;  however in scintillator and argon detectors, there could in principle be ``IceCube style" detection of \cevns interactions.  IceCube, which has very sparse photomultiplier arrays, does not detect individual IBD interactions; rather, it collects single photons from the diffuse glow of supernova-neutrino-induced Cherenkov photons in the ice~\cite{Halzen:1994xe}.  Similarly, single photons from the diffuse glow of \cevns interactions from scintillation in liquid hydrocarbon or argon could be collected over the time scale of the burst.  The back-of-the-envelope calculation is as follows:  there are about two orders of magnitude more \cevns than CC interactions in a given target, but about three orders of magnitude less energy deposition per interaction.  Furthermore, there is typically a quenching factor of at least a few in photon production for recoils of heavy particles (nuclei) with respect to light ones (electrons or positrons).  On the other hand, there is a factor of six for \cevns with respect to CC, given that the neutrino flux is approximately equally divided among flavors.  Overall this results in few to ten percent of \cevns-induced photons with respect to CC-induced photons.  However the \cevns glow photons should be diffused over the burst rather than in short, inelastic-interaction-associated spikes.

The primary issue for detection of \cevns glow is background.  Preliminary calculations show that cosmogenic $^{39}$Ar $\beta$ decays, which have a rate of about a Bq/kg in natural argon,  may completely swamp the signal in argon.  Underground argon, depleted in this isotope, could mitigate this.\footnote{Note that there can also be ``\cevns buzz" from ionization collected in liquid argon TPCs.}  Large liquid scintillator detectors are likely quieter.  One will also know the time frame of the burst given inelastic event detection.  The distribution of photon numbers as a function of time may be a handle for extracting signal from background. This idea is ambitious and it is not yet clear it is feasible, but we are continuing to study it.

%
% image bounding boxes
%
% figure_scint-glow
% 0 0 794 237
%
% figure_argon-glow
% 0 0 793 240
%
%
\begin{figure}[h]
\centering
\includegraphics[width=6in, bb = 0 0 793 240]{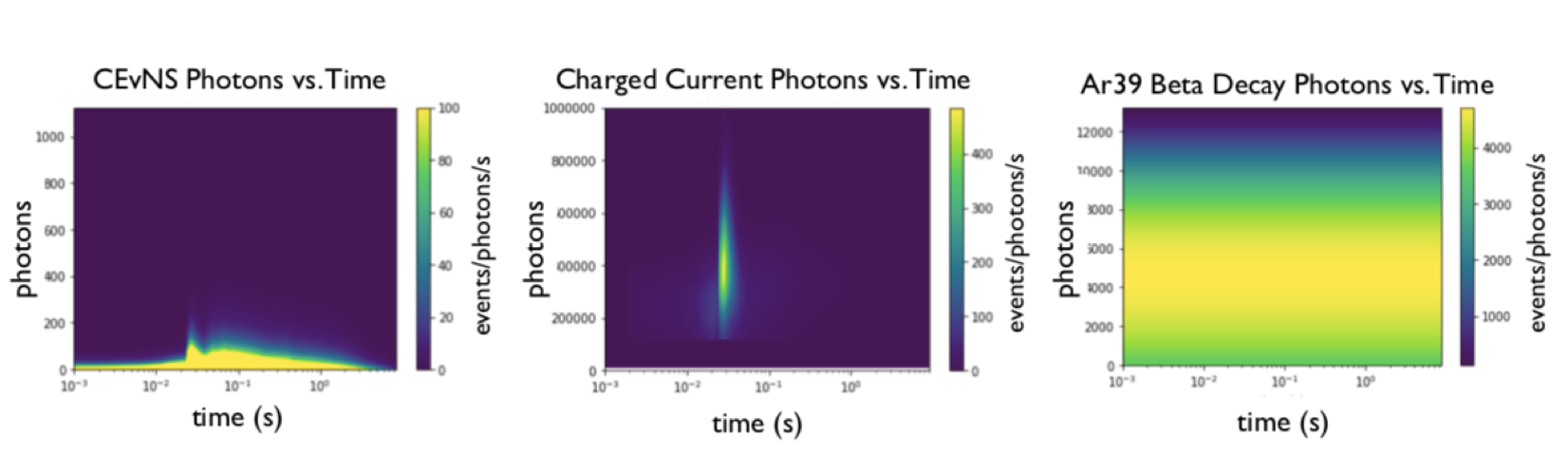}
\includegraphics[width=6in, bb = 0 0 794 237]{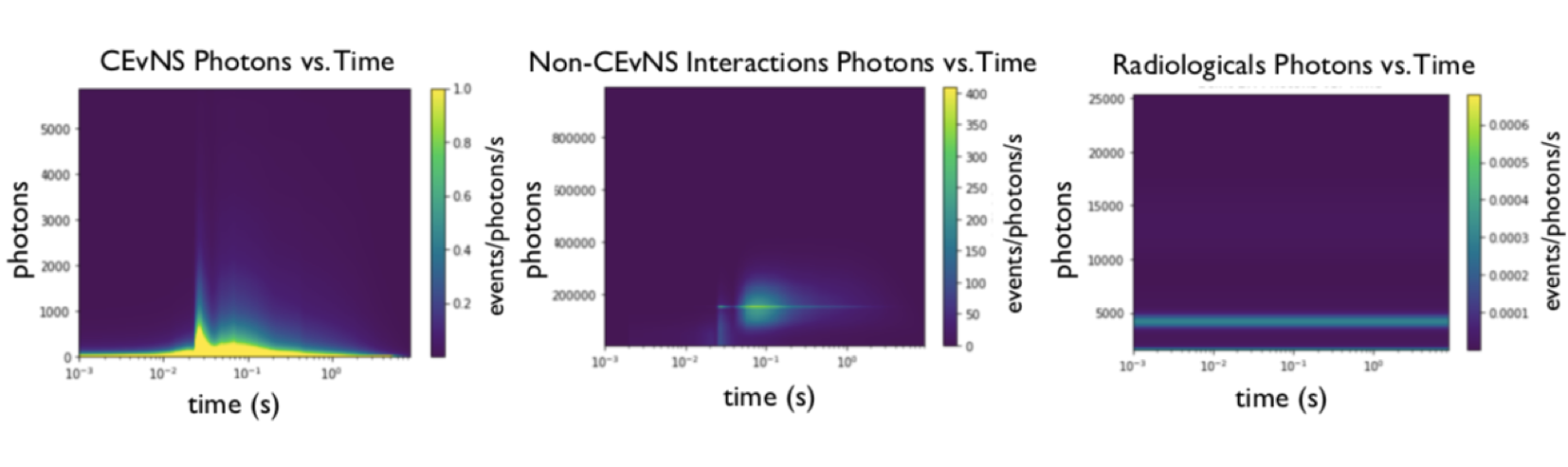}
\caption{Distributions of numbers of photons arriving at a detection surface as a function of time, for the supernova neutrino production model of Ref.~\protect{\cite{Huedepohl:2009wh}}, from a preliminary analytic calculation by A. Smith. Top plots: argon scintillation.  Bottom plots: liquid hydrocarbon scintillation.  The left plots show photons from \protect{\cevns}; center plots are photons from the primary charged- and neutral-current channels in the respective material, and right plots show photons due to backgrounds. }
\label{fig:cevnsglow}
\end{figure}

\chapter{\cevns as a Probe of Nuclear Neutron Form Factors}
\chapterauthor{Kelly Patton}

\chapteraffil{Institute for Nuclear Theory, University of Washington, Seattle, WA 98195, USA}

\chaptercoauthor{G. McLaughlin, K. Scholberg, J. Engel, N. Schunck}

\chapterdoi{10.5281/zenodo.3464614}

The size of a nucleus is one of its most fundamental properties.  While the distribution of protons has been well measured through the use of charged probes, the neutron distribution has remained difficult to probe.  Previous measurements have used hadronic scattering and report uncertainties on the root-mean-square (RMS) radius of $\sim1\%$ \cite{Gibbs:1992lt, Trzcinska:2001zy}, but require assumptions about the underlying nuclear structure to obtain results.  The PREX experiment used parity violating electron scattering to measure the neutron skin and RMS radius of $^{208}$Pb, reporting an uncertainty of $2.5\%$ on the RMS radius \cite{ref:prexI}.  CE$\nu$NS provides a new method for measuring the RMS radius for a range of nuclei.

All nuclear structure information in the CE$\nu$NS cross section is included in the form factor $F(Q^{2})$, which is defined as

\begin{equation}
F(Q^{2}) = \frac{1}{Q_{W}} \int{(\rho_{n}(r) - (1-4\sin^{2}\theta_{W})\rho_{p}(r)) \frac{\sin(Qr)}{Qr}r^{2}dr}.
\end{equation}
Here, $Q_{W}$ is the weak charge of the nucleus, and $\rho_{n,p}$ are the neutron and proton densities.  Since this process is very low energy, we can Taylor expand the form factor and write it in terms of moments of the distribution \cite{Patton:2012jr}.  So, for example, the neutron terms can be written as
\begin{equation}
F_{n}(Q^{2}) \approx N \left( 1 - \frac{Q^{2}}{3!} \langle R_{n}^{2} \rangle + \frac{Q^{4}}{5!} \langle R_{n}^{4} \rangle + \cdots \right),
\end{equation}
where $\langle R_{n}^{k} \rangle$ is the $k^{th}$ moment of the distribution.  Changes in the values of these moments has an effect on the number of events as a function of energy, which can be measured in a CE$\nu$NS experiment. For a detector placed 20 m from the source at the SNS, corresponding to the location of Neutrino Alley, we find that the RMS radius can be measured with an uncertainty of $\sim5\%$ for Ar, Ge, and Xe, using detectors of 3.5 tonnes, 1.5 tonnes, and 300 kg, respectively.  These calculations assumed an detector threshold of 5 keV.  It is also possible to get the first experimental measurement of the fourth moment of the neutron distribution, with uncertainties on the order of $\sim10-20\%$ \cite{Patton:2012jr}.  These measurements also depend on detector uncertainties, in particular the spectral shape uncertainty which describes the difference in efficiency between energy bins.  We have found that this specific uncertainty must be known to the level of $\sim1\%$ to measure the RMS radius to $5\%$ at the $90\%$ confidence level for Ar, Ge, and Xe \cite{Patton:2013}.  These results show that CE$\nu$NS is a promising new method of probing neutron distributions in nuclei.  
\chapter{Future sensitivity of \cevns to a weak mixing angle}
\chapterauthor{Omar Miranda}

\chapteraffil{Departamento de F\'isica, Centro de Investigaci\'on y de Estudios Avanzados del IPN, Apdo. Postal 14-740, 07000 Ciudad de M\'exico, M\'exico}

 \chapterdoi{10.5281/zenodo.3464592}

Precise measurements of \cevns will allow to measure the weak mixing
angle with precision. 
There is room of improvement for this important parameter of the standar model in the low energy
regime~\cite{Tanabashi:2018oca}, where \cevns measurements are
performed. We have studied the potential of future \cevns detectors
located close to reactor antineutrino fluxes~\cite{Canas:2018rng}.
For this kind of measurements, the ratio of protons to neutrons is
very relevant since the dependence on the weak mixing angle is present
only in the proton coupling constants. 
Besides, for large statistic experiments, the reactor antineutrino flux uncertainties will also be
an important issue to solve, since current uncertainties translate
into a systematic error around one percent for the weak mixing
angle measurement. 
\chapter{Neutrino constraints on conventional and exotic \cevns interactions}
\chapterauthor{D.K. Papoulias}

\chapteraffil{Instituto de F\'isica Corpuscular (CSIC-Universitat de Val\`encia), Paterna (Valencia), Spain}

\chaptercoauthor{T.S. Kosmas}

\chapterdoi{10.5281/zenodo.3464608}

\begin{table}[t]
%\begin{tabular}{cc}
\begin{tabularx}{\linewidth}{X@{\extracolsep{\fill}}c}
\toprule
%\toprule
Parameter & Limit (90\% C.L.) \\
\midrule
$\sin^2 \theta_W$ & 0.117 -- 0.315\\
%\hline
$\epsilon_{ee}^{uV}$ & -0.08 -- 0.47 \\
$\epsilon_{ee}^{dV}$ & -0.07 -- 0.42 \\
$\epsilon_{\mu \mu }^{uV}$ & -0.09 -- 0.48 \\
$\epsilon_{\mu \mu}^{dV}$ & -0.08 -- 0.43 \\
%\hline
$\epsilon_{ee}^{uT}$ & -0.013 -- 0.013 \\
$\epsilon_{ee}^{dT}$ & -0.011 -- 0.011 \\
$\epsilon_{\mu \mu }^{uT}$ & -0.013 -- 0.013 \\
$\epsilon_{\mu \mu}^{dT}$ & -0.011 -- 0.011 \\
%\hline
$\mu_\nu$ & 4.3 $\times 10^{-9} \mu_B$ \\
$\mu_{\nu_e}$ & 5.2 $\times 10^{-9} \mu_B$\\
$\mu_{\nu_\mu}$ & 4.6 $\times 10^{-9} \mu_B$\\
%\hline 
$\langle r^2_\nu \rangle$ & -31.4 -- -23.1 and -4.9 -- 3.4 \\
$\langle r^2_{\nu_e} \rangle$ & -38.0 -- -26.6 and -1.4 -- 10.1 \\
$\langle r^2_{\nu_\mu} \rangle$ & -39.6 -- -27.4 and -0.6 -- 11.7\\
\bottomrule
%\bottomrule
\end{tabularx}
\caption{Constraints on SM and exotic parameters from the first observation of CE$\nu$NS at the COHERENT experiment. For the case of the neutrino charge radius the results are given in units of $10^{-32} \mathrm{cm^2}$~\cite{Kosmas:2017tsq}.}
\label{papoulias:table:summary}
\end{table}
Assuming one non-zero parameter at a time, the current extracted 90\% C.L. limits on the weak mixing angle, NSI and electromagnetic properties (neutrino magnetic moment and charge radius) are summarized in Table~\ref{papoulias:table:summary}~\cite{Kosmas:2017tsq}. The next phase of the COHERENT program, on the basis of a multi-target strategy with ton-scale detectors, will lead to significant improvements on the current determination of the weak mixing angle as well as to improvements on NSI, sterile neutrino and new mediator (vector $Z^\prime$ and scalar $\Phi$) constraints by one order of magnitude (see e.g.~\cite{Papoulias:2015vxa,Kosmas:2017zbh}). Next generation advances of CE$\nu$NS experiments with ultra low-threshold technologies are expected to provide severe constraints on neutrino electromagnetic properties that will compete with current neutrino-electron scattering data~\cite{Kosmas:2015sqa,Kosmas:2015vsa}. The latter will also enable validation of the neutrino-floor and detector-response models relevant to Dark Matter searches~\cite{Papoulias:2018uzy}.

\chapter{Aspects of Elastic Scattering of Neutrinos}
\chapterauthor{A.B. Balantekin}

\chapteraffil{Department of Physics, University of Wisconsin, Madison, WI 53706, USA}

\chapterdoi{10.5281/zenodo.3463600}

Coherent elastic scattering of neutrinos off nuclei was experimentally observed for the first time only recently \cite{coherentScience2017}. The associated cross section is sensitive to the neutron distribution in the target nuclei. Even a single extra neutron appreciably increases the coherent scattering cross section as illustrated in Ref. \cite{Suzuki:2019cra}
for \iso{12}{C} versus \iso{13}{C}. 

Elastic scattering of neutrinos off nuclei is primarily governed by weak neutral current scattering. Since neutrinos have mass we know that neutrino magnetic moments are non-zero, but we do not know just how 
large they are \cite{Giunti:2014ixa,Balantekin:2018azf}. 
Neutrino elastic scattering can also have a small electromagnetic component due to  
neutrino magnetic moments. The differential scattering cross section is then sum of the two components: 
\begin{eqnarray}
\frac{d\sigma}{dT} &=& \frac{G_F^2}{8 \pi} Q_W^2 M \left[ 2 - 2 \left( \frac{T}{T_{\rm max}} \right) + \left( \frac{T}{E} \right)^2  \right] \left[ F_Z(Q^2)\right]^2 \nonumber \\
&+& \frac{\pi \alpha^2}{m_e^2} Z^2 \mu_{\rm eff}^2 \left[ \frac{1}{T} - \frac{1}{E} \right] \left[ F_{\gamma} (Q^2)\right]^2 \nonumber
\end{eqnarray}
where $E$ is the neutrino energy, $T$ and $M$ are the recoil energy and the mass of the nucleus, respectively, and $Q_W = [N-(1-4\sin^2 \theta_W)]$. The effective neutrino magnetic moment is given by 
\begin{equation}
 \mu_{\rm eff}^2 = \sum_i \left| \sum_j   U_{({\rm e \> or} \> \mu)j}  e^{-i E_jL} \mu_{ji} \right|^2  \nonumber
\end{equation}
where $L$ is the distance traveled by the neutrino before it scatters, $E_j$ is the energy of the $j$-th mass eigenstate, and $\mu_{ij}$ are the elements of the neutrino magnetic moment matrix in the mass basis. Note that not only the second term in the above cross section is smaller than the first term, but two form factors $F_Z(Q^2)$ and $F_{\gamma} (Q^2)$ represent represent two rather different aspects of the nuclear structure, i.e. primarily neutron versus proton distributions in the nuclei. 

\chapter{Measurement of \cevns with LAr}
\chapterauthor{Rex Tayloe}

\chapteraffil{Indiana University, Bloomington, IN 47405, USA}

\chapterdoi{10.5281/zenodo.3464652}

The ORNL Spallation Neutron Source provides the world's most intense pulsed neutron beams.  The 1.4~MW, 1.0~GeV pulsed proton beam also provides a world-class pion decay-at-rest neutrino source with a 600~ns width, 60~Hz repetition rate for a duty-factor of $10^{-4}$.  This beam, combined with a low-energy-threshold LAr detector, will allow for low-background, precision studies of the coherent elastic neutrino nucleus scattering (\cevns) process.  An initial run with the CENNS-10 22~kg fiducial volume detector has demonstrated that a low energy threshold and small backgrounds can be obtained in a large volume.   Design parameters for the next-generation COHERENT LAr detector are:
\begin{itemize}
\item detector mass (total/fiducial): 750(612)~kg
 \item integrated beam power (protons-on-target):  7000~MWhr/yr (1.6E23POT/yr)
 \item  pion decays-at-rest/protons-on-target:  0.9
 \item pion production target - detector distance: 27.5~mag
 \item light yield: $5\pm1$~photoelectrons/keVee
 \item argon quenching factor: $0.29\pm3\%$
 \item energy threshold: 20~keVnr
 \item estimated detected \cevns event sample:  3000 events/year
\item  signal/background ratio with atmospheric (underground \cite{undergroundArgonWorkshop}) argon:  1:10 (1:1) 
\end{itemize}
\chapter{A Search for \cevns with the CENNS-10 Liquid Argon Detector for COHERENT}
\chapterauthor{Jacob Zettlemoyer}

\chapteraffil{Indiana University, Bloomington, IN 47405, USA}

\chapterdoi{10.5281/zenodo.3464707}

The COHERENT Collaboration deploys the CENNS-10 detector, a 22-kg liquid argon detector, at the Spallation Neutron Source at Oak Ridge National Laboratory for a measurement of \cevns on a light nucleus to complement other planned measurements within COHERENT. 
The detector began operation in December 2016 and was upgraded for improved light collection in July 2017 and has been running since the upgrade. 
The initial run will be used to measure the flux of neutrons that occur in time with the SNS beam pulse at the CENNS-10 location and confirm previous measurements. 
With the full shielding structure, simulations show the main component of the steady-state background is the internal component \iso{39}{Ar}. 
Using the timing of the SNS beam and the pulse shape discrimination capabilities of liquid argon, these backgrounds can be reduced to levels necessary for a \cevns measurement. 
The analysis of the physics data is ongoing. 
The current parameters are:
\begin{itemize}
    \item Detector mass: 22 kg, fiducial.
    \item Shielding: 20 cm cylindrical water tank, 1.27 cm copper on all sides outside water tank, and 10.16 cm lead on all sides outside of the copper.
    \item Light Yield: 4.2 $\pm$ 0.2 photoelectrons/keVee.
    \item Threshold: 20 keVnr.
    \item Steady state backgrounds: measured 1.7 Hz in energy ROI (0 -- 35 keVee), includes SNS timing.
    \item Energy Resolution: 9.1$\%$ at 41 keVee.
    \item Quenching factor: $\textrm{QF} = aE + b$ with $a = 0.251$ and $b = 7.52 \times 10^{-4}$ with $E$ in keVnr from 0-120 keVnr.
\end{itemize}
\chapter{Spherical proportional counters and their application for CEnNS detection}
\chapterauthor{Marie Vidal}

\chapteraffil{Department of Physics, Engineering Physics \& Astronomy, Queen's University, Kingston, Ontario K7L 3N6, Canada}

\chaptercollab{NEWS-G}

\chapterdoi{10.5281/zenodo.3464675}

The NEWS-G collaboration uses spherical proportional counters to search for low mass Dark Matter. The important features and results of the experiment are listed below:
\begin{itemize}
\item Flexibility in gas choice: noble gases and operating pressure
\item Detector sensitivity to single electron: low energy threshold
\item Constraints in the Spin-Independent WIMP-nucleon cross section vs WIMP mass: 0.6 GeV. For LSM data (2017): 9.6 kg.days with a mixture of Neon + 0.7\% CH$_4$: \cite{arnaud2018news}
\item Preliminary projection for next experiment (2019): below 0.1 GeV. For a Neon + 10\% CH$_4$ and sensitivity down to $\sim$ 10$^{-41}$ cm$^2$.
\end{itemize}
Here are some preliminary calculation for CEnNS detection using reactor neutrino, considering:
\begin{itemize}
\item 1 GW thermal power
\item 10m from core
\item Neutrino flux: Vogel
\end{itemize}
Detector conditions:
\begin{itemize}
\item Low threshold 100 eV$_{ee}$ or below achieved
\item Moderate size sphere: 80 cm of diameter
\item 2 bar of Argon: $\sim$ 1kg
\end{itemize}
Including detector response:
\begin{itemize}
\item Mean energy to create e$^-$/ion pair: 36 eV in Neon: \cite{inokuti1975,combecher1980}
\item quenching factor: Lindhard parametrization
\item Poisson distribution
\item single electron response: \cite{brossard2018}
\end{itemize}
Event rate:
\begin{itemize}
\item Total event rate: 112 events/kg/day
\item Event rate above 50 eV$_{ee}$: 28 events/kg/day
\item Event rate above 100 eV$_{ee}$: 15 events/kg/day
\end{itemize}
\chapter{Progress on liquid-noble bubble chambers for \cevns}
\chapterauthor{C. Eric Dahl}

\chapteraffil{Northwestern University, Evanston, IL 60208, USA}

\chaptercollab{SBC}

\chapterdoi{10.5281/zenodo.3464442}

\section*{Detector Concept}

\subsection*{Scintillating argon bubble chamber}
Nuclear recoils (NRs) from \cevns in a superheated liquid argon target can both produce scintillation light and nucleate a single bubble in the superheated fluid.  Electron-recoil (ER) backgrounds (beta-decays, gammas) produce scintillation light only.  This does-it-make-a-bubble discrimination is effective at much lower thresholds than pulse-shape discrimination in the scintillation signal --- we expect $10^{-8}$ ER sensitivity at sub-keV NR bubble nucleation thresholds.

\begin{figure}[!h]
\begin{center}
\includegraphics[width=4.5in, trim = 0 0 0 0, clip = true, bb=0 0 468 216]{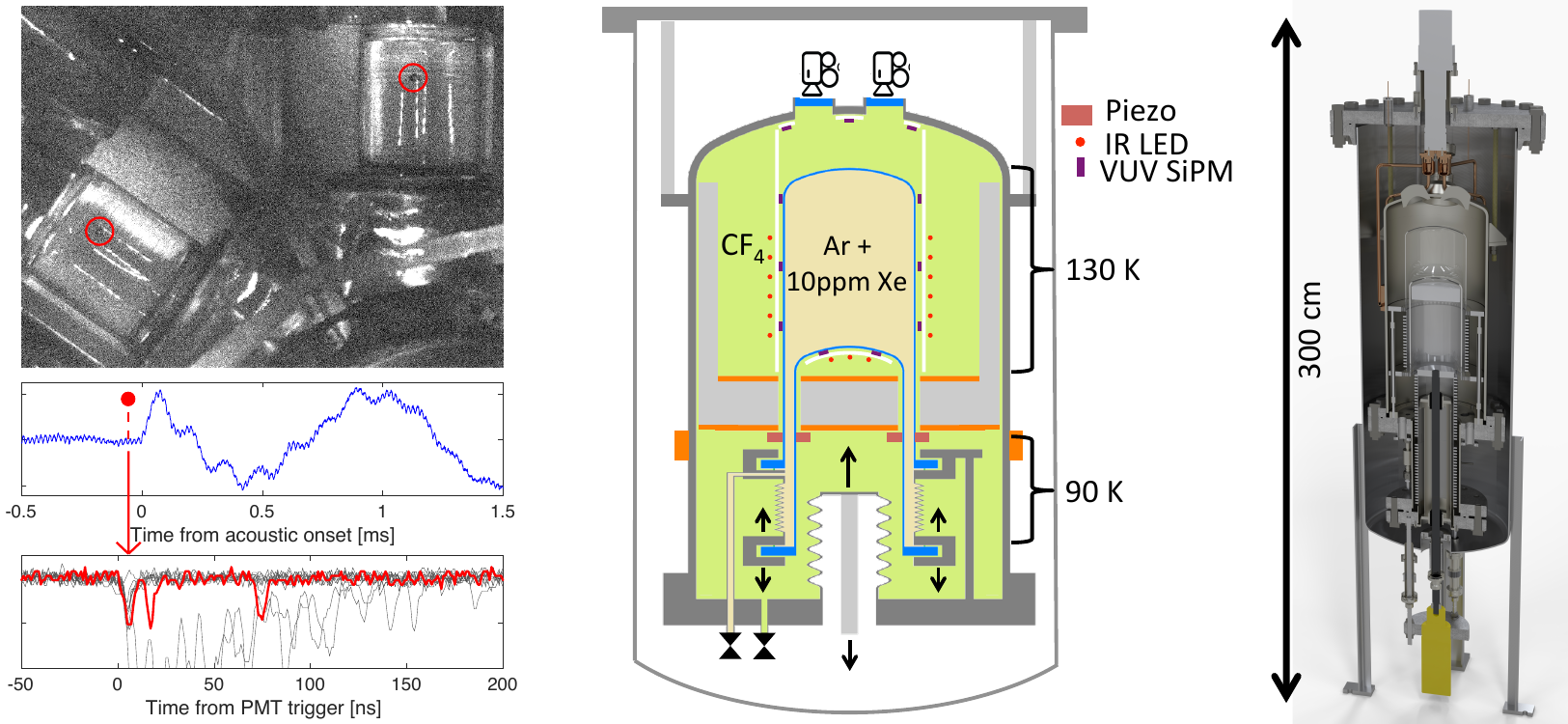}
\caption{\label{F-DM-xebc}\emph{Left:} Sample nuclear recoil event from the prototype xenon bubble chamber \cite{PhysRevLett.118.231301}.  Top:  Stereo image of a single xenon vapor bubble.  Middle: Acoustic record (blue) of bubble formation, giving the time of nucleation to $\pm$25~$\mu$s.  In this case nucleation is coincident with a scintillation trigger (red). The lag between the scintillation pulse and acoustic onset matches the sound speed in liquid xenon.  Bottom:  PMT waveforms showing xenon scintillation. The bubble-coincident pulse is shown in red.  ER's generate scintillation pulses without coincident bubble nucleation (gray traces).  \emph{Left:} Schematic and model of the 10-kg argon bubble chamber, showing the pressure and temperature control, bubble imaging, and scintillation detection scheme for the chamber.  The solid model and rendering were done by the FNAL PPD/Mech Eng Dept.}
\end{center}
\end{figure}

\section*{Key Performance Specifications: Goals for the Fermilab Scintillating Bubble Chamber}

{\renewcommand{\arraystretch}{1.4}
\begin{tabular}{|l|lp{3in}|}
    \hline
    {\bf Target Material/Mass} & --- & 10-kg superheated argon @ 25 psia, 130 K \\
    \hline
     {\bf Bubble threshold} & --- & Effecient NR bubble nucleation @ 0.1 keV \\
      & --- & $10^{-8}$ ER sensitivity \\
     \hline
     {\bf Scintillation threshold} & --- & 1 photon detected / 2 keV (NR) \\
      & --- & Bubble-nucleating events with 0 photons detected are included in the \cevns signal region \\
      \hline
      {\bf Energy Resolution} & --- & Event-by-event energy resolution from scintillation:  2~keV \\ 
       & --- & Spectral resolution from bubble nucleation threshold scan:  0.1~keV \\
       \hline
       {\bf Spatial resolution} & --- &  1-mm absolute resolution, from stereo imaging of bubble \\
        & --- &  0.1-mm resolution for indentifying multi-scatter events, from acoustic signal strength \\
        \hline
        {\bf Timing resolution} & --- & 10~ns for events with coincident scintillation signal \\
        & --- & 25~$\mu$s for events without coincident scintillation ($\lesssim$2~keV NR) \\
        \hline
        {\bf Backgrounds} & --- & Negligible ER backgrounds, including \iso{39}{Ar}, due to bubble discrimination \\
        & --- & No high-Z shielding required, reducing neutrino-induced-neutron backgrounds \\
        & --- & Signifcant dead time ($\sim$30 seconds) after every bubble-nucleating event limits the background NR rate to $<$0.1~Hz.  Operations require overburden and/or water/poly shield. \\
        \hline
\end{tabular}
}
\chapter{LArCADe: lowering thresholds in LArTPC detectors}
\chapterauthor{David Caratelli}

\chapteraffil{Fermi National Accelerator Laboratory, Batavia, IL 60510, USA}

\chaptercoauthor{A. Fava}

\chapterdoi{10.5281/zenodo.3464434}

The recent detection of \cevns by the COHERENT collaboration, and the rich physics program that CEvNS can deliver, motivate the optimization of existing detectors and development of new technologies for the purpose of improving the sensitivity to the experimental signature of coherent neutrino scattering processes. 
The LArCADe project aims to investigate the feasibility of reducing detection thresholds for ionization electrons in single-phase Liquid Argon Time Projection Chamber (LArTPC) detectors by enough to enable the detection of nuclear recoils. 
The program aims to allow for the amplification of drifting electron signals directly in the liquid phase by modifying the geometry of the charge-collecting anode sensors. 
Such a technological achievement could merge into one the advantages of kiloton-scale liquid argon detectors and those of low-threshold double-phase dark-matter TPCs.
	
Nuclear recoils in liquid argon lead to small ionization signals, further reduced by the significant quenching caused by ion recombination and dissipation of energy into atomic excitations. 
Nuclear recoils of 1~--~10s of keV, originating from $\mathcal{O} \! \left(10~\text{MeV}\right)$ \cevns interactions, are expected to yield 1~--~100 free electrons, with significant variation in the tails of such distributions depending on the assumed quenching model \cite{scene2015,agnes2017a,agnes2018a}.
These values are a factor of 100 smaller than current state of the art detection thresholds in LArTPCs \cite{adams2018a}. 
In order to amplify ionization charge directly in the liquid phase, strong fields of $ >\!10^5$  V/cm are necessary. 
The LArCADe program is exploring the possibility of obtaining stable charge amplification of drifting electrons by shaping the electric field over micron-scale distances in the proximity of the charge-collecting anode-planes. 
The first phase of this R\&D effort is employing tungsten tips of micron radii, and has demonstrated preliminary controlled amplification in gaseous argon using  a few-cm drift chamber which records ionization charge produced by a pulsed LED source impinging on a photocathode. 
A second phase, currently underway, aims to use $\mathcal{O} \! \left(100~\text{nm}\right)$ tips to obtain amplification in liquid, characterizing stability and potential complications which may arise, such as the formation of argon gas bubbles which can disrupt signal detection. 
A successful demonstration of this program can lead in the future to the construction of small-scale detectors sensitive to \cevns interactions in the proximity of intense neutrino beams.
\chapter{Dark side of the exciton: self-organized criticality and low energy threshold detectors}
\chapterauthor{Sergey Pereverzev}

\chapteraffil{Lawrence Livermore National Laboratory, Livermore, CA 94550, USA}

\chaptercoauthor{A. Bernstein, J. Xu}

\chapterdoi{10.5281/zenodo.3464619}

With interest to detect coherent scattering of low energy solar and reactor neutrinos on nuclei of the detector materials, we analyses low energy response of different detectors. 
It is common for ionization and scintillation detectors to demonstrate increase of low-energy background at energies of the order of excitations in the detector material- i.e., at the level of few electrons or photons.  
Practically in all solid materials, including rare gases solid,  slow irradiation  by muons and  residual radioactivity leads to  accumulation of energy  in form of trapped charges (pairs of ions, trapped electrons and holes)- which lead to effects as thermally-stimulated luminescence, thermally stimulated exaelectron emission (electron emission from the surface of dielectrics or dielectric films on metal surfaces), thermally stimulated conductivity (in semiconductors and dielectrics). 
Defects / impurities clustering is another common effect in solids. 
Thus, one can expect to see Self-Organized Criticality type of dynamics- slow accumulation of excitations and events of their annihilation in form of small avalanches. 
In dual-phase Ar and Xe detectors solid phase is present in a form of solid physiosorbed films on all internal surfaces. 
Native positive ions Xe${}_{2^+}$ (Ar${}_{2^+}$) , and negatively charged complexes formed out Xe (Ar) and O, F, H atoms  (these can be due to small residual amount of  oxygen, water,  and fluoride compounds coming from  TEFLON) can be produced due to ionization events and due to exposure of solid films on surfaces to  UV light (gas electroluminescence  light- S2 pulses).  
Accumulation of those species on electrodes and their avalanche-like annihilation can lead to ``few electrons events'' which mimic real low energy particles detection events. 
By application of strong AC electric field in between grid wires ion recombination on the cathode grid can be reinforced and ions accumulation suppressed. 
Decrease of parasitic background in this case can be experimentally verified. 

Our analysis illustrates that searches for rare and low energy particles interactions require careful examination of the detector physics and advanced studies of condensed matter effects.

This work was performed under the auspices of the U.S. Department of Energy by Lawrence Livermore National Laboratory under Contract DE-AC52-07NA27344.
\chapter{The development of low threshold dual phase argon detector in China for \cevns measurement}

\chapterauthor{Ran Han}

\chapteraffil{Beijing Institute of Spacecraft Environment Engineering, Beijing 100094, China}

\chaptercoauthor{W. Yang}

\chapterdoi{10.5281/zenodo.3464505}

\begin{figure}
\begin{center}
\includegraphics[width=0.5\textwidth, bb = 0 0 1101 948]{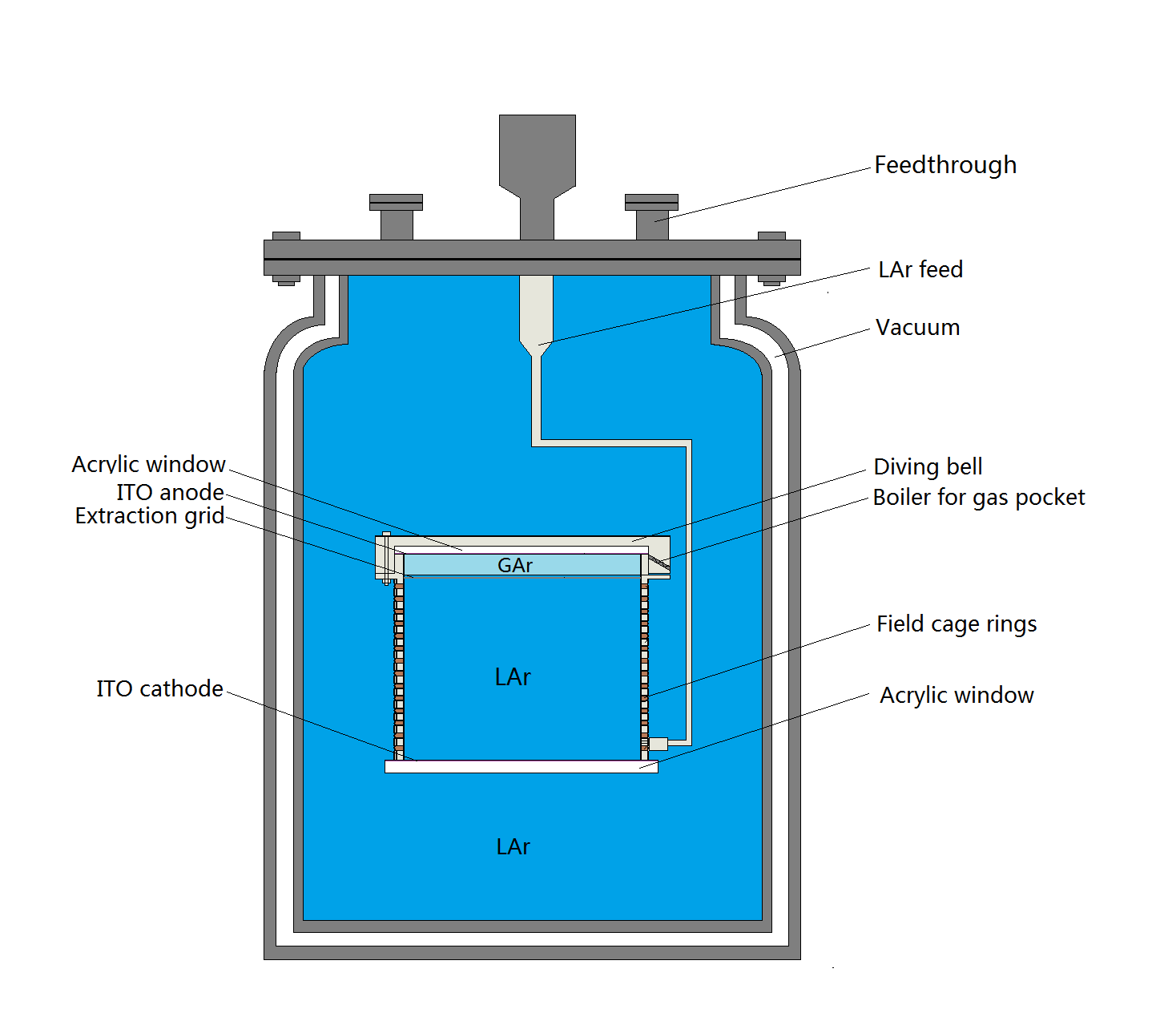}
\end{center}
\caption{The design drawing of the LAr detector.} \label{fig:hanran1}
\end{figure}

The dual-phase liquid argon time projection chamber (TPC) is designed for the coherent elastic neutrino-nucleus scattering \cite{freedman74,freedman77} research. 
The detector are planned to settled near the core of Taishan reactor (located in Guangdong province) at the distance of 31 m.
The power of the Taishan reactor is 4.6 GW which provides a reactor flux around $4.37 \times 10^{13} / \left(\text{cm}^2 \cdot \text{s}\right)$ in total.
The TPC is designed as a cylinder while the dimensions of TPC are: the diameter $d = 56~\text{cm}$, the height $h = 58~\text{cm}$ and the mass of liquid argon is about 200 kg, see Fig. \ref{fig:hanran1}.
Three electrodes are used to generate the drift field, extraction field and collection field, respectively.
Gas pocket is generated by the vaporization of the liquid argon which thickness is 10 mm.
The grid is placed below the gas-liquid interface 5 mm.
Drift field is design as 400 V/cm after simulation.
The inner container is made of acrylic material to reduce the radio activities background.
SiPMs are used to collect S1 and S2 signals instead of photomultiplier tubes (PMTs). 
Properties of SiPM and acquisition systems at low temperature are under study. 
To obtain the low threshold results, only S2 signals are collected.

%
% bounding box for image
%
% 0 0 1101 948
%

To detect the low recoil energy in the dual-phase liquid argon TPC, quenching factor have to be precise determined. 
Quenching factor defined as the ratio of the light yields of the nuclear recoils with respect to the ones of the electron recoils at the same energy. 
Generally, the electron recoils are taken as no quenching effect and set to be unit. 
In order to determine the quenching factor, nuclear recoil energies should be measured. 
Nuclear recoil energies or deposit energies can be measured according to the following equation \cite{arneodo2000}
\begin{equation*}
	E_r \approx 2 E_n\frac{M_\text{n} M_\text{Ar}}{\left(M_\text{n} + M_\text{Ar}\right)^2} \left( 1 - \cos{\theta} \right),
\end{equation*}
where $E_r$, $E_n$ are the nuclear recoil energy and the neutron beam energy, respectively. 
$M_\text{n}$ is the mass of a neutron and $M_\text{Ar}$ is the mass of the argon nucleus.
$\theta$ is the scattering angle of the outgoing neutron.
To reduce the background and the systematic errors, the TPC have to be the smaller the better.
The diameter of the sensitive region is designed as 7.6 cm while the sensitive mass is about 0.5 kg.
Neutron detectors are placed at relatively large distance from the TPC for the small scattering angle neutron detection.
Neutron detectors arrays are also used to increase the statistics at different angles, respectively. 
Shield can be used to separate two kinds of neutron beams, coming directly from the neutron source beam and the scattering neutron beam, for small scattering angle detections.
For the measurement of quenching factor at sub-keV scale, sub-MeV neutron source have to be used and the TPC also have to be exposed for a few mouths to obtain enough data.

%%%%%%%%%%%%%%%%%%%%%%%%%%%%%%%%%%%%%%%%%%%
% bibliographic section
% \bibliographystyle{utphys}
% \bibliography{magCEvNS2018}

%\begingroup
%\titlespacing{\chapter}{0pt}{-2\baselineskip}{\baselineskip}
%{\singlespacing
%\printbibliography[heading=bibintoc,title={BIBLIOGRAPHY}]}
%\endgroup

\printbibliography[heading=bibintoc,title={REFERENCES}]

\end{document}